\documentclass[fleqn,usenatbib]{mnras}
\pdfoutput=1

\usepackage{newtxtext,newtxmath}

\usepackage[T1]{fontenc}

\DeclareRobustCommand{\VAN}[3]{#2}
\let\VANthebibliography\thebibliography
\def\thebibliography{\DeclareRobustCommand{\VAN}[3]{##3}\VANthebibliography}

\newcommand{\msun}{{\rm M}_{\odot}}

\usepackage{graphicx}	
\usepackage{amsmath}	
\usepackage{subcaption}
\usepackage{xcolor}
\usepackage{orcidlink}

\title[GOGREEN \& simulations]{GOGREEN: a critical assessment of environmental trends in cosmological hydrodynamical simulations at $z\approx1$}

\author[Kukstas et al.]{Egidijus Kukstas$^{1,2,3}$\orcidlink{0000-0001-6902-501X},
Michael L. Balogh$^{4,5}$\thanks{mbalogh@uwaterloo.ca}\orcidlink{0000-0003-4849-9536},
Ian G.~McCarthy$^{1}$\thanks{i.g.mccarthy@ljmu.ac.uk}\orcidlink{0000-0002-1286-483X},
Yannick M.~Bah{\'e}$^{6}$\orcidlink{0000-0002-3196-5126},
\newauthor
Gabriella De Lucia,$^{7}$\orcidlink{0000-0002-6220-9104}
Pascale Jablonka$^{8,9}$\orcidlink{0000-0002-9655-1063},
Benedetta Vulcani$^{10}$\orcidlink{0000-0003-0980-1499},
Devontae C. Baxter$^{11}$\orcidlink{0000-0002-8209-2783},
\newauthor
Andrea Biviano$^{7,12}$\orcidlink{0000-0002-0857-0732},
Pierluigi Cerulo$^{13}$\orcidlink{0000-0003-0703-3123},
Jeffrey C. Chan$^{14}$\orcidlink{0000-0001-6251-3125},
M. C. Cooper$^{11}$\orcidlink{0000-0003-1371-6019},
\newauthor
Ricardo Demarco$^{15}$\orcidlink{0000-0003-3921-2177},
Alexis Finoguenov$^{16}$\orcidlink{0000-0002-4606-5403},
Andreea S. Font$^{1}$\orcidlink{0000-0001-8405-9883},
Chris Lidman$^{17,18}$\orcidlink{0000-0003-1731-0497},
\newauthor
Justin Marchioni$^{4}$,
Sean McGee$^{19}$\orcidlink{0000-0003-3255-3139},
Adam Muzzin$^{20}$\orcidlink{0000-0002-9330-9108},
Julie Nantais$^{21}$\orcidlink{0000-0002-7356-0629},
Lyndsay Old$^{22}$\orcidlink{0000-0003-1205-1318}, 
\newauthor
Irene Pintos-Castro$^{23}$\orcidlink{0000-0002-9133-4457},
Bianca Poggianti$^{10}$\orcidlink{0000-0001-8751-8360},
Andrew M. M. Reeves$^{4,5}$\orcidlink{0000-0003-2618-6408},
Gregory Rudnick$^{24}$\orcidlink{0000-0001-5851-1856},
\newauthor
Florian Sarron$^{25}$,
Remco van der Burg$^{26}$\orcidlink{0000-0003-1535-2327},
Kristi Webb$^{4,5}$\orcidlink{0000-0002-8610-0672},
Gillian Wilson$^{14}$\orcidlink{0000-0002-6572-7089},
\newauthor
Howard K. C. Yee$^{27}$\orcidlink{0000-0003-4935-2720},
Dennis Zaritsky$^{28}$\orcidlink{0000-0002-5177-727X}
\\
Author affiliations are listed at the end of the paper
}

\date{Accepted XXX. Received YYY; in original form ZZZ}

\pubyear{2022}

\begin{document}
\label{firstpage}
\pagerange{\pageref{firstpage}--\pageref{lastpage}}
\maketitle

\begin{abstract}
Recent observations have shown that the environmental quenching of galaxies at $z \sim 1$ is qualitatively different to that in the local Universe.  However, the physical origin of these differences has not yet been elucidated.  In addition, while low-redshift comparisons between observed environmental trends and the predictions of cosmological hydrodynamical simulations are now routine, there have been relatively few comparisons at higher redshifts to date.  Here we confront three state-of-the-art suites of simulations (BAHAMAS+MACSIS, EAGLE+Hydrangea, IllustrisTNG) with state-of-the-art observations of the field and cluster environments from the COSMOS/UltraVISTA and GOGREEN surveys, respectively, at $z \sim 1$ to assess the realism of the simulations and gain insight into the evolution of environmental quenching.  We show that while the simulations generally reproduce the stellar content and the stellar mass functions of quiescent and star-forming galaxies in the field, all the simulations struggle to capture the observed quenching of satellites in the cluster environment, in that they are overly efficient at quenching low-mass satellites.  Furthermore, two of the suites do not sufficiently quench the highest-mass galaxies in clusters, perhaps a result of insufficient feedback from AGN.  The origin of the discrepancy at low stellar masses ($M_* \la 10^{10}$ M$_\odot$), which is present in all the simulations in spite of large differences in resolution, feedback implementations, and hydrodynamical solvers, is unclear.  The next generation of simulations, which will push to significantly higher resolution and also include explicit modelling of the cold interstellar medium, may help to shed light on the low-mass tension.
\end{abstract}




\section{Introduction}

Observational galaxy surveys of large, statistical samples have demonstrated that the population of gas-poor galaxies with negligible star formation rates built up gradually over time \citep[e.g.][]{Bell_etal_2004,Faber_etal_2007,2007ApJ...654..858B,Muzzin_etal_2013} as star formation activity ended.  The bimodality in the distributions of colours and star formation rates out to $z\approx 2$ \citep[e.g.][]{Strateva_etal_2001, Blanton_etal_2003, Baldry_etal_2004,Kauffmann_etal_2004,Noeske_etal_2007, Gallazzi_etal_2008, Brammer_etal_2009, Whitaker_etal_2011,Muzzin_etal_2013} suggests that this cessation of star formation must occur fairly rapidly, in a process dubbed ``quenching''.  The precise physical causes for this transition are unknown, but likely depend on the galaxy mass, epoch and large-scale environment.

\citet{Peng_etal_2010} have demonstrated that, at least at low redshift, the fraction of quenched galaxies depends on both stellar mass and environment in a way that is separable \citep[see also][]{Baldry_etal_2006}.  It has been hypothesized that this represents distinct physical processes: internal (e.g., stellar and AGN feedback) driving the stellar mass dependence, and external (e.g., tidal or ram pressure stripping of gas) responsible for the environmental trends. 

Observations at higher redshift, however, are revealing a more complex picture, with multiple studies finding correlations that are very different from what is observed at $z=0$ \citep{Balogh_etal_2016, Kawinwanichakij_etal_2017, Papovich_etal_2018, vdBurg_etal_2018, Pintos-Castro_etal_2019}.  In particular, in contrast to the findings of \citet{Peng_etal_2010}, the correlations with environment do not appear to be independent of stellar mass \citep[e.g.][]{Balogh_etal_2016,Kawinwanichakij_etal_2017}.  Analysis of galaxy clusters at $1<z<1.4$ from the GOGREEN \citep{Balogh_etal_2017,GOGREEN_DR} survey suggests in fact that the physics behind quenching the most massive cluster galaxies $\mathrm{log_{10}(M_*/M_{\odot}) > 10.5}$ in dense environments may be significantly different from that affecting the less massive galaxies \citep[see also ][]{Poggianti06}.  The ages of the most massive cluster galaxies suggest that they ceased forming stars at $z>2$, likely before they were ever part of a massive, virialized halo \citep{Webb}.  This is further supported by the observations that such galaxies are already quenched in much lower mass haloes \citep{Reeves}, and in the distant infall regions of the GOGREEN clusters \citep{Werner}.  The implication is that whatever quenched star formation happened at high redshift, in the moderately overdense regions around protoclusters.  The apparent environment independence of the quenched galaxy stellar mass function (GSMF) suggests that the quenching mechanism of such galaxies may be independent of environment, but simply happens in an accelerated, or earlier, fashion in protoclusters \citep{vdBurg_etal_2020}.  On the other hand,  lower mass quiescent galaxies in clusters appear to be consistent with a scenario in which they all quenched recently, upon infall into the main cluster progenitor \citep[see also][]{Kawinwanichakij_etal_2017}.  This is supported by the abundance of post-starburst galaxies \citep{McNab}, and the fact that the quenched fraction of such galaxies is negligible in groups \citep{Reeves} and the infall regions \citep{Werner}.  In that case, their quenching is likely related directly to the environment.

While the general role of feedback in quenching central galaxies is reasonably well established, there is no consensus yet on exactly how the environment increases the quenching rate for satellite galaxies, leading to an excess quenched fraction.  There are many proposed processes, including hydrodynamic interactions between gas in the galaxy and its host such as ram-pressure stripping \citep{Gunn_Gott_1972,Quilis_etal_2000,Barsanti_etal_2018} or strangulation/starvation \citep{Larson_etal_1980, Balogh_etal_2000, McCarthy_etal_2008, Peng_etal_2015}, or purely  gravitational interactions such as galaxy-galaxy mergers \citep{Mihos_Hernquist_a_1994, Mihos_Hernquist_b_1994, Schawinski_etal_2014} and harassment \citep{Farouki_Shapiro_1981, Moore_etal_1999, Hirschmann_etal_2014}.  However, sophisticated models implementing these processes still generally fail to reproduce in detail the observed fraction of star-forming galaxies in clusters, or its stellar mass dependence \citep[e.g.][]{2008MNRAS.389.1619F,2012MNRAS.426.2797W}.  Interestingly, the failure is usually in the sense that the models predict a fraction of quenched galaxies in clusters that is too high, indicating that the environmental dependence is too strong/efficient in the models.  Continued development of physically motivated, but ad-hoc, recipes in some semi-analytic models has generally led to improvements, where in some cases the match to $z=0$ data is reasonably good \citep[e.g.][]{2020MNRAS.498.4327X}.

Cosmological hydrodynamical simulations have made great strides in the last decade, such that they can now self-consistently solve the equations for cosmic evolution starting from Gaussian perturbations in the density field and culminating with present-day, realistic-looking galaxies embedded within voids, sheets, filaments, and clusters (e.g., \citealt{Schaye_etal_2015, McCarthy_etal_2017, Pillepich_etal_2018}). 
Once the subgrid models for feedback processes in such simulations are calibrated to reproduce certain global properties of the galaxy population, such as the local GSMF, they can make useful independent predictions for the environmental dependence of galaxy properties \citep{Bahe_McCarthy_2015}.  This is because the physical processes associated with environment (including assembly history, tidal disruption and hydrodynamic interactions) are calculated self-consistently.  Unlike in simple toy or more complex semi-analytic models, subgrid approximations are not needed to represent these effects (although how effective they are may depend on numerical resolution).  The effectiveness of these environmental processes can nevertheless be sensitive to the modelling of subgrid processes such as feedback, as both the properties of the satellites and the host are altered when the feedback is altered.  For example, if the feedback is too strong and results in low gas densities in the simulated galaxies, this could lead to the galaxies being overly susceptible to ram pressure stripping when they fall into a galaxy cluster.  Therefore, careful comparison between data and models with different subgrid implementations can shed light on the realism of these prescriptions.  Simulations also allow the detailed study of environmental processes, beyond what can be directly observed, to aid our understanding of the origin of model successes and failures.   

The most recent and successful cosmological hydrodynamical simulations are calibrated to reproduce some subset of observational properties typically dominated by field galaxies (e.g., the GSMF).  
At present, most comparisons between these simulations and observations of galaxies in the densest regions of the Universe (galaxy clusters) have been limited to low redshifts, where large, complete redshift survey data exists. Such comparisons have shown some success in matching the observed demographics of cluster galaxies  \citep[e.g.][]{Donnari_etal_2021}, though in many cases they
still over-predict the quenched fraction at low masses  (e.g., \citealt{Bahe_etal_2017,Lotz_etal_2019}).
With the recent completion of the GOGREEN survey at a higher redshift of $z\approx 1$, 
it is timely to make a careful comparison of those data with state-of-the-art simulations. 

The structure of this paper is as follows. In Sect.~\ref{sec:observations} we give a brief description of the data from the GOGREEN survey.  In Section~\ref{sec:sims} we describe the three main suites of cosmological hydrodynamical simulations used in this study.  In Sect.~\ref{sec:results} we perform a detailed comparison of the simulations with GOGREEN data, focusing on the quenched fraction and its dependence on mass and environment.  Finally, in Sect.~\ref{sec:conclusions} we summarise the findings and present the conclusions drawn.  Note that all simulations and observations used in the present study adopt a \citet{Chabrier_2003} stellar initial mass function (IMF).

\section{Observational Data}
\label{sec:observations}
The observations of galaxies in rich clusters are taken from the GOGREEN \citep{Balogh_etal_2017,GOGREEN_DR} cluster survey, which consists of homogeneous, deep imaging and spectroscopy of 21 galaxy groups and clusters at $1<z<1.5$. For this study we restrict the comparison to the ten massive clusters at $z<1<1.4$ (mean redshift $z=1.23$) analysed in \citet{vdBurg_etal_2020}.   Halo masses have been determined from Jeans modeling of the redshift distribution, as summarized in \citet{GOGREEN_DR} and described in detail in \citet{Biviano_etal_2021}.  They span a range of $10^{14.1}< M_{200{\rm c}}/\msun < 10^{14.8}$, with a mean of $10^{14.5}~\msun$.   The halo mass distribution of these ten clusters are shown in Figure~\ref{fig:hmf} as the broken green histogram.  
Both observations \citep[e.g.][]{Reeves} and many previous simulation-based studies have shown that properties such as the quenched fraction and quenched fraction excess can depend on halo mass (e.g., \citealt{Bahe_etal_2017_hydrangea,Donnari_etal_2019,Donnari_etal_2021}).  It will therefore be important to consider this distribution when comparing the data with simulations, below.

Stellar masses are computed from the total observed {\it K}-band magnitudes, with a mass-to-light ratio determined from template fitting to the spectral energy distributions assuming a Chabrier IMF \citep{Chabrier_2003}.  The GSMFs presented in \citet{vdBurg_etal_2020} are determined using both spectroscopic and photometric redshifts and are statistically complete for stellar masses of $M_* \gtrsim 10^{9.5}~\msun$. 

A comparison field galaxy sample is taken from the COSMOS/UltraVISTA DR1 \citep{Muzzin_etal_2013} catalogue.  This consists of galaxies from a 1.69~$\mathrm{deg^2}$ field, with photometric redshifts in the range of $1.0 < z < 1.4$, complete down to stellar mass of $10^{9.5}~\msun$. Completeness corrections have been applied as per Sec. 4.2 of \citet{vdBurg_etal_2020}.

Following common convention, both cluster and field galaxies were classified by \citet{vdBurg_etal_2020} as star-forming and quiescent based on their rest-frame $U$-$V$ and $V$-$J$ colours \citep{W+09,Muzzin_etal_2013}.  The galaxy distribution shows a distinct bimodality in this plane that allows the two populations to be separated in a way that is not strongly dependent on dust or metallicity.   This is discussed further, below, in Section \ref{sec-systematics}. 

\section{Simulations}
\label{sec:sims}
We consider results from three broad suites of simulations:  BAHAMAS (BAryons and Haloes of MAssive Systems, \citealt{McCarthy_etal_2017,McCarthy_etal_2018}); EAGLE (Evolution and Assembly of GaLaxies and their Environments; \citealt{Schaye_etal_2015, Crain_etal_2015, McAlpine_etal_2016}); and the TNG300 simulation, part of the IllustrisTNG project \citep{Pillepich_etal_2018,Springel_etal_2018}.  For BAHAMAS and EAGLE we also consider an associated set of re-simulations of massive haloes, using the same physics as the parent simulation: MACSIS (MAssive ClusterS and Intercluster Structures, \citealt{Barnes_etal_2017}) and Hydrangea \citep{Bahe_etal_2017}, respectively.  
The simulation box sizes and particle masses (mass resolution) are summarised in Table~\ref{table:sims}.  The most relevant characteristics of these simulations (including the calibration of their subgrid feedback parameters, and key distinguishing features that directly impact the predictions considered in this paper) will be summarized in more detail in the following subsections.  

\begin{table}
\begin{centering}
\begin{tabular}{ccccc}
\multicolumn{1}{c}{\textbf{Name}} & \multicolumn{1}{c}{\textbf{L (cMpc)}} & \multicolumn{1}{c}{\textbf{N}} & \multicolumn{1}{c}{$\boldsymbol{\mathrm{m_b\ (M_{\odot})}}$} & \multicolumn{1}{c}{$\boldsymbol{\mathrm{m_{DM}\  (M_{\odot})}}$} \\ \hline
BAHAMAS                             & $571$                                  & $2\times 1024^3$                               & $1.09\times 10^9$                                         & $5.5\times 10^9$                                           \\
EAGLE                               & $50$                                   & $2\times 752^3$                                & $1.81\times 10^6$                                         & $9.7\times 10^6$                                            \\
TNG300-1                              & $303$                                  & $2\times 2500^3$                               & $1.1\times 10^7$                                          & $5.9\times 10^7$                                           
\end{tabular}
\caption[A summary of the properties of the simulation used.]{A comparison of the periodic-box simulations used in this study: length of the cubic simulation box in co-moving megaparsecs, number of particles (baryonic and dark matter), and mass of each type of particle. The zoom-in simulations, MACSIS and Hydrangea, use the same resolution parameters as their corresponding periodic box simulations, BAHAMAS and EAGLE, respectively. Particle masses have been converted to physical units using the appropriate value for $h$ where necessary.}
\label{table:sims}
\end{centering}
\end{table}

Note that all of the analysis presented in this paper is done using catalogue-level data, rather than working directly with the particle data.

\subsection{Simulation codes and parameters}
\subsubsection{BAHAMAS and MACSIS}
\label{sec:BAHAMAS_MACSIS}

The BAHAMAS \citep{McCarthy_etal_2017,McCarthy_etal_2018} project is a set of smoothed particle hydrodynamic (SPH) simulations carried out using a significantly modified version of Gadget-3 (last described by \citealt{Springel_2005}) and available in a variety of different cosmologies. In this study, we make use of the fiducial simulation presented in \citet{McCarthy_etal_2017}, which adopts the WMAP 9-year maximum-likelihood cosmology \citep{Hinshaw_etal_2013}.  The simulations were performed in a periodic cube of length $\mathrm{L}=596\ \mathrm{cMpc}$ and $2\times 1024^3$ particles with masses of $\approx5.5\times 10^9~\msun$ and $\approx1.09\times 10^9~\msun$ for dark matter and baryons, respectively.  Note that, if the conditions for star formation are satisfied, a single gas particle is converted into a single star particle.  The star particle can then lose mass due to stellar evolution, transferring some of its mass (and metals) to neighbouring gas particles.  Thus, the mass of gas particles is not precisely preserved during the simulation and can (typically) vary by up to a factor of 2 from the initial mass.

A number of subgrid physics models, originally developed for the OWLS project \citep{Schaye_etal_2010}, are used for physics which cannot be resolved directly in the simulations.  Specifically, radiative cooling rates are computed on an element-by-element basis by interpolating within pre-computed tables generated with CLOUDY, that contain cooling rates as a function of density, temperature and redshift calculated in the presence of the CMB and photoionization from a \citet{Haardt_2001} ionizing ultraviolet/X-ray background (see \citealt{Wiersma_etal_2009_39399}). Star formation is tracked in the simulations following the prescription of \citet{Schaye_Dalla_Vecchia_2008}. Gas with densities exceeding the critical density for the onset of the thermogravitational instability is expected to be multiphase and to form stars \citep{Schaye_2004}.  Since the simulations lack both the physics and the resolution to model the cold interstellar gas phase, an effective equation of state (EOS) is imposed with pressure $P \propto \rho^{4/3}$ for densities $n_{\rm H} > 0.1$ cm$^{-3}$.  Gas on the effective EOS is allowed to form stars at a pressure-dependent rate that reproduces the observed Kennicutt–Schmidt law by construction. The timed release of individual elements (‘metals’) by
both massive (SNe II and stellar winds) and intermediate-mass stars (SNe Ia and AGB stars) is included following the prescription of \citet{Wiersma_etal_2009_399574}. A set of 11 individual elements are followed in these simulations (H, He, C, Ca, N, O, Ne, Mg, S, Si and Fe), which represent all the important species for computing radiative cooling rates.  For a more complete description of the above, the reader is referred to \citet{Schaye_etal_2010}.

The parameters characterising the efficiencies of the stellar and AGN feedback were adjusted to reproduce the observed GSMF and the amplitude of group/cluster gas mass fraction--halo mass relation (as inferred from resolved X-ray observations) at $\mathrm{z\approx 0}$. The aim of the calibration was to ensure the simulations have the correct total baryon content in collapsed haloes, so that the simulations realistically capture the effects of baryons on the matter power spectrum \citep{van_Daalen_etal_2020}, which is the basis of most large-scale structure tests of cosmology.  Stellar feedback is implemented using the isotropic kinetic model of \citealt{Dalla_Vecchia_Schaye_2008}, where neighbouring gas particles are given a velocity `kick'.  The number of gas particles (the mass-loading) and the velocity kick are the free parameters which are varied to reproduce the low-mass end of the present-day GSMF.  AGN feedback is implemented using the isotropic thermal model of \citet{Booth_Schaye_2009}, where selected neighbouring gas particles have their temperatures boosted by a certain amount, $\Delta T_{\rm heat}$.  The number of gas particles selected for heating and the temperature jump are free parameters which were varied to roughly reproduce the knee of the observed GSMF and the gas fractions of galaxy groups and clusters, respectively.  Note that the BH sink particles store accretion energy (accreted at a rate proportional to the local Bondi-Hoyle-Lyttleton rate and which is Eddington-limited) until there is sufficient energy to heat the specified number of particles by the chosen value of $\Delta T_{\rm heat}$.  Thus, increasing $\Delta T_{\rm heat}$ results in more energetic but also more bursty (less frequent) feedback episodes.  A detailed discussion of the calibration of the feedback models is presented in \citet{McCarthy_etal_2017}.  

MACSIS \citep{Barnes_etal_2017} is an ensemble of 390 `zoom-in' simulations centred on individual haloes drawn from a 3.2~Gpc $N$-body simulation. These re-simulations use the same code as BAHAMAS with the same subgrid prescriptions and parameter values, and were run at the same resolution as outlined above, resulting in a set of massive haloes which ideally supplements the sample of haloes available from the main BAHAMAS box.

The BAHAMAS model has been demonstrated to reproduce reasonably well the evolution of the GSMF for $z\lesssim 2.5$ \citep{McCarthy_etal_2017} and the local cluster X-ray and Sunyaev-Zel'dovich effect scaling relations \citep{Barnes_etal_2017}. However, it appears to over-quench low-mass galaxies in high-density regions in the local Universe \citep{Kukstas_etal_2020}. We will revisit this point later.

\subsubsection{EAGLE and Hydrangea}
\label{sec:EAGLE_Hydrangea}

The EAGLE \citep{Schaye_etal_2015, Crain_etal_2015, McAlpine_etal_2016} simulation is also run with a version of Gadget-3.  However, the hydrodynamics solver differs from that used for BAHAMAS (which used the solver of \citealt{Springel_Hernquist_2003}), in that it uses the pressure-entropy SPH formalism of \citet{Hopkins_2013}, artificial viscosity switch \citep{Cullen_Dehnen_2010}, artificial conductivity switch \citep{Price_2008}, and time-step limiter of \citet{Durier_Dalla_Vecchia_2012}.  These changes, collectively referred to as `Anarchy SPH', are not expected to be important at the resolution of BAHAMAS but become more important at higher resolutions.

EAGLE uses a very similar set of subgrid physics models as BAHAMAS, as both are descendant from the OWLS project.  Aside from minor updates to the cooling rates and details of the EOS implementation, perhaps the most significant differences with respect to BAHAMAS are in the stellar feedback (which is implemented thermally in EAGLE, as opposed to kinetically in BAHAMAS) and the way stellar and AGN feedback were calibrated.  Specifically, the stellar feedback parameters in EAGLE were adjusted to reproduce the global GSMF and the sizes of galaxies at $z\approx0$.  Because the stellar feedback parameters in both BAHAMAS and EAGLE were adjusted to reproduce the GSMF, we do not anticipate differences in the nature of the feedback (kinetic vs. thermal) between the simulations to be significant, at least for the stellar content.  No specific calibration was made for the AGN feedback parameters in the fiducial EAGLE `Reference' model, however, and comparisons with X-ray observations of galaxy groups reveal that it predicts gas fractions in excess of those observed.  We, therefore, use the `AGNdT9' model, which was run after the large Reference simulation, in a smaller box.  As the name suggests, the AGN subgrid heating temperature was adjusted (raised) with respect to the Reference model, in order to better provide an improved match to the observed gas fraction--halo mass relation of groups (although it still predicts gas fractions that are somewhat too high for the most massive systems), as well as the observed X-ray luminosity-temperature relation, while retaining a good fit to the local GSMF and galaxy sizes (see \citealt{Schaye_etal_2015} for details). In addition, the black hole subgrid accretion disc viscosity parameter $C_\mathrm{visc}$ was increased from $2\upi$ to $200\upi$. It is also the model adopted in the Hydrangea zooms described below. 

Note that all EAGLE simulations adopt cosmological parameters from \citet{Planck2013_2014}, which are slightly different from the WMAP 9-year values adopted for BAHAMAS.  Notably, the values of $\Omega_m$ and $\sigma_8$ preferred by \textit{Planck} ($0.307$ and $0.829$, respectively) are larger than those preferred by the WMAP data ($0.279$ and $0.821$, respectively) and also more in tension with large-scale structure constraints (e.g., \citealt{McCarthy_etal_2018}).  However, for the purposes of environmental studies, we do not expect such differences in the cosmological parameter values to be important\footnote{Indeed, \citet{McCarthy_etal_2018} have compared the resulting galaxy and cluster properties (scaling relations, etc.) in WMAP and Planck cosmologies, finding near identical results.}.  We will refer to AGNdT9 model simply as `EAGLE' throughout this paper, as it is the only variant we use. 

The EAGLE AGNdT9 model was run in a 50~cMpc periodic box (with $N=756^3$ and $\mathrm{m_{baryon}} = 1.81\times 10^6~\msun$ and $\mathrm{m_{DM}} = 9.7\times 10^6~\msun$), meaning that very few haloes above $M_{200{\rm c}} \approx 10^{14.0}~\msun$ exist. For this reason, we supplement the sample with the Hydrangea \citep{Bahe_etal_2017} suite of zoom-in re-simulations.  Importantly, Hydrangea uses the same galaxy formation model and resolution as AGNdT9 run described above, allowing for the two to be combined seamlessly.  These haloes were selected from the same 3.2~Gpc $N$-body simulation as MACSIS haloes, although there is no overlap in the specific haloes identified for the two projects.

\begin{figure*}
	\includegraphics[width=\textwidth]{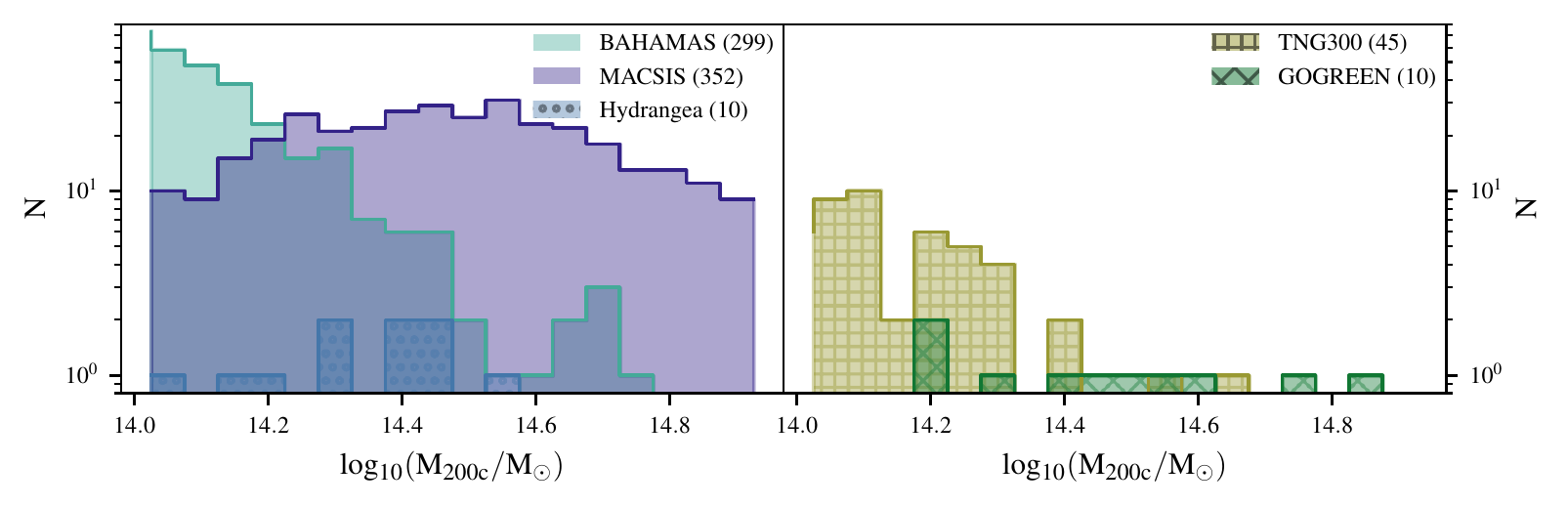}
    \caption{Halo mass distributions for all simulations from which clusters have been selected, as well as for GOGREEN (shown in green). Numbers in the legend indicate the number of haloes in each sample. The periodic box-based simulations (BAHAMAS, TNG300) are generally not sufficiently large enough to fully sample the GOGREEN cluster halo mass range. In the case of EAGLE, no haloes exist at this halo mass range. Supplementing BAHAMAS with MACSIS and (to an extent) EAGLE with Hydrangea helps to resolve this problem. The figure is split into two panels purely for visual clarity.}
    \label{fig:hmf}
\end{figure*}

\subsubsection{Illustris TNG300}
\label{sec:TNG300}

The TNG300 simulation, part of the IllustrisTNG project \citep{Pillepich_etal_2018,Springel_etal_2018}, uses a very different type of hydrodynamic solver to those described above.   Specifically, it uses the `moving mesh' magneto-hydrodynamics \citep{Pakmor_etal_2011, Pakmor_Springel_2013} and gravity solver, AREPO \citep{Springel_2010}.  In addition to this, it implements subgrid physical models for metal-dependent radiative cooling, star formation and a simple multiphase treatment of the interstellar medium, stellar population evolution and chemical synthesis, stellar feedback, and the formation and feedback mechanisms of supermassive black holes. 

The subgrid models of IllustrisTNG include similar physics as for EAGLE, although the implementation often differs significantly. Radiative cooling rates are computed based on total metallicity using the ionizing background model from \citet{Faucher-Giguere_et_al_2009}, with self-shielding in the dense ISM explicitly taken into account. The ISM is modelled using the two-phase model of \citet{Springel_Hernquist_2003} with an effective EOS. As in EAGLE, stars are formed stochastically in gas that exceeds a density threshold of $n_\mathrm{H} \gtrsim 0.1$ cm$^{-3}$, based on the Kennicutt-Schmidt relation and assuming a \citet{Chabrier_2003} initial mass function. Energy feedback from star formation, sourced by star forming gas rather than stochastically formed star particles, is injected in kinetic form, with a velocity that is explicitly scaled with the local DM velocity dispersion and redshift, and a metallicity-dependent mass loading factor \citep{Pillepich_etal_2018_methods}; stellar winds are temporarily decoupled from the hydrodynamics. For AGN feedback, the simulations use the two-mode model of \citet{Weinberger_etal_2017} with energy injected in an energetically inefficient thermal mode at high Eddington fractions, and an energetically efficient kinetic mode at low accretion rates. The transition between the two regimes depends on BH mass. In contrast to stellar feedback, winds driven by AGN are not kinematically decoupled. For an in-depth description of the models, the interested reader is referred to \citet{Pillepich_etal_2018_methods} and \citet{Weinberger_etal_2017}. IllustrisTNG adopts cosmological parameters consistent with \citet{Planck2015_2016}.

The TNG model was calibrated to reproduce several observed trends, including the present-day GSMF, black hole mass--stellar mass relation, galaxy size--stellar mass relation, and gas fractions of low-mass galaxy groups.  The simulation parameters were also adjusted to better reproduce the observed evolution of the cosmic star formation rate density. While attempts were made to calibrate TNG on group gas-fractions, this was done in relatively small calibration volumes.   The most massive clusters in TNG300 have somewhat too high gas fractions \citep{Barnes_etal_2019}.
The standard TNG300(-1) simulation has $2\times 2500^3$ particles, with $\mathrm{m_{baryon}} = 1.1\times 10^7~\msun$ and $\mathrm{m_{DM}} = 5.9\times 10^7~\msun$. 

We note that for the TNG300 simulation (which is lower resolution than the calibrated TNG100 simulation), there are actually two flavours of catalogues: one using the stellar masses, star formation rates, etc. predicted directly by the simulation and another where quantities have been rescaled to better agree with the higher-resolution TNG100 simulation (often denoted `rTNG').  Note that it is TNG100 which was explicitly calibrated against observations as described above, whereas TNG300 is the same model run in a larger volume at lower resolution.  We use the unscaled TNG300 predictions but we have verified that using the rTNG variant does not significantly affect our results or conclusions. 

\subsection{Halo and galaxy selection}
\label{sec:halo_selection_fq_desig}

All simulations used in this study identify dark matter haloes in a common way.   Specifically, a standard 3D friends-of-friends (FOF)  finder (with a linking length set to 0.2 of the mean interparticle separation) is first run on the dark matter particles to identify FOF groups.  Gas, star, and BH particles in close spatial proximity to the FOF DM particles are then attached to the FOF group.  Following this, the \verb|SUBFIND| algorithm \citep{Springel_etal_2001, Dolag_etal_2009} is run on all particle types to identify self-gravitating substructures.  Particles are initially assigned to potential subhaloes by looking for overdense structures.  An energy unbinding procedure is then used to identify which particles (if any) in the overdensity are truly bound (i.e., part of a subhalo).  Aside from a minimum number of particle constraint to be deemed a subhalo (20 particles), there is no constraint on the content of a subhalo.  For example, subhaloes can in principle be composed entirely of star particles or DM particles.  For the stellar mass cuts we adopt for the simulations when comparing to GOGREEN (see below), however, virtually all of our subhaloes have stars (by construction), DM, BHs, and often some gas (aside from those that have been completely ram pressure stripped).

For all simulations, we consistently measure the stellar mass as that which is bound to a subhalo and within a spherical 30 kpc (physical) aperture.  \citet{Schaye_etal_2015}, \citet{McCarthy_etal_2017}, and \citet{Pillepich_etal_2018} have previously shown for EAGLE (and therefore also Hydrangea), BAHAMAS (and therefore MACSIS), and TNG300, respectively, that a 30 kpc aperture is a good approximation for standard observational pipeline-based (e.g., Petrosian) stellar masses, which do not include the contribution from diffuse intracluster light.  For consistency, we also measure star formation rates within the same aperture, although we comment below on the impact of changing this aperture.  
In the case of TNG300, 30 kpc aperture measurements were not available for the star formation rates. Instead, an aperture equal to twice the stellar half-mass radius, $\mathrm{R_{*,h}}$, was used as the nearest equivalent.  We do not expect this difference in SFR aperture to be significant, though, since the SFR is generally a centrally-concentrated quantity.

In EAGLE/Hydrangea and TNG300, only simulated galaxies with $\mathrm{log_{10}(M_{*}/M_{\odot}) > 9.0}$ are included in the analysis.  This is mainly driven by the fact that the GOGREEN sample is stellar mass complete to $\mathrm{log_{10}(M_{*}/M_{\odot}) > 9.5}$ with which we later compare (described below). In the case of BAHAMAS/MACSIS, the resolution limits the stellar mass range to $\mathrm{log_{10}(M_{*}/M_{\odot}) > 10}$. Nevertheless, a comparison can still be made between all simulations over the majority of GOGREEN sample range.

The closest common snapshot redshift for all five simulations is $z=1.0$. This is somewhat lower than the observations (which have a mean $z=1.23$).  This means the simulated galaxy abundance at fixed stellar mass will be slightly higher than it would for a sample better matched to the data.   
However, the difference is smaller than either the error bars on data points or the $1\sigma$ variance in the simulated cluster population.  An approximate magnitude of this difference can be seen in figure~13 of \citet{McCarthy_etal_2017} where GSMFs for $z=1.0$ and $z=1.5$ are plotted in the top right panel.  We therefore ignore the small degree of evolution that is expected to occur between $z=1.5$ and $z=1.0$.

In Figure~\ref{fig:hmf} we show the $z \approx 1$ distribution of halo masses in each simulation considered in this paper, and compare with the GOGREEN clusters.  Owing to its large periodic volume (400 Mpc/$h$ on a side), BAHAMAS has the largest number of haloes (299) with $M_{200{\rm c}} > 10^{14}~\msun$.  The other periodic volumes, namely TNG300 (45) and EAGLE, have significantly fewer systems (albeit at significantly higher resolution), with the 50 Mpc EAGLE volume having no systems above this mass limit at $z=1$.  Consequently, the EAGLE volume will only be used for computing `field' properties for the combined EAGLE/Hydrangea analysis.  As noted above, we supplement the high-mass end of BAHAMAS with the MACSIS zoom simulation suite and do likewise for EAGLE with the Hydrangea suite.  TNG300 does not, at present, have an accompanying suite of zooms.

\begin{figure*}
	\includegraphics[width=0.8\textwidth]{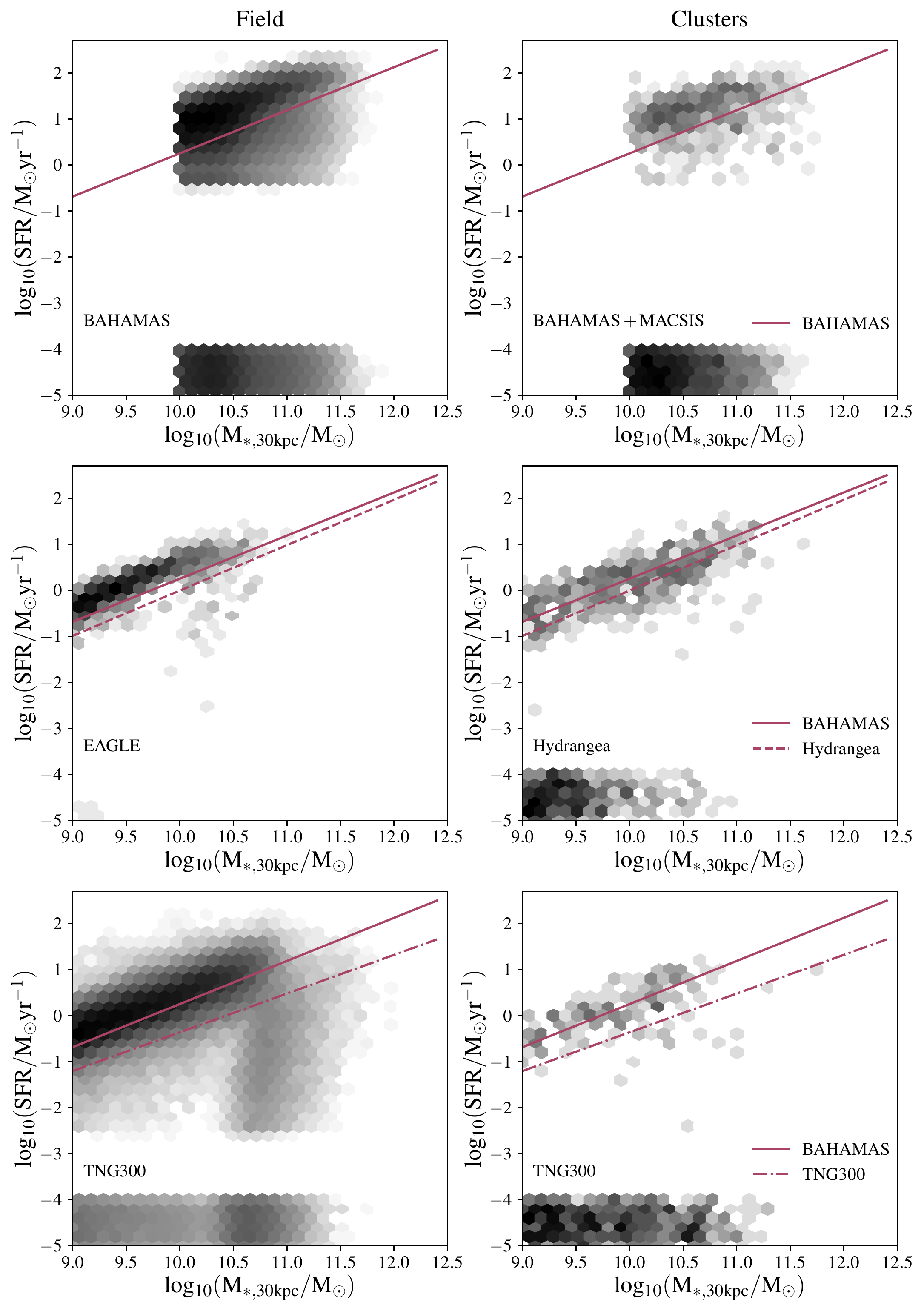}\\
    \caption{ Field and cluster distributions of $\mathrm{log_{10}(SFR) - log_{10}(M_{*})}$ for all three sets of simulations used in this study. Diagonal lines indicate the star-forming--quiescent division as determined using the SFMS-fitting method (see text). The solid line for BAHAMAS is repeated in all panels for comparison with the other simulations.
    }
    \label{fig:SFR_dist}
\end{figure*}

To make a fair comparison between the simulations and GOGREEN, we need to ensure a reasonable correspondence in the cluster halo mass distributions of the simulations.  For BAHAMAS/MACSIS, there are sufficiently numerous high-mass haloes that we can draw many independent samples of ten GOGREEN-like haloes. This presents an opportunity to explore the possible scatter in the GSMF estimate due to cosmic variance. For Hydrangea and TNG300, however, the most massive haloes are still not massive enough to properly match those present in GOGREEN. For Hydrangea, there is no simple alternative but to select the ten most massive haloes shown in Figure~\ref{fig:hmf}; these have a mean $M_{200{\rm c}} = 10^{14.34}~\msun$.  TNG300 suffers from the same lack of massive haloes at high-mass end but contains a much higher number of low-mass systems. We therefore select ten
unique haloes from TNG300 matching the GOGREEN halo mass distribution as closely as possible.  This sample has a mean halo mass of $M_{200{\rm c}} = 10^{14.36}~\msun$.  

\subsection{Star formation rates and galaxy classification}\label{sec:sims-sfr}

In Figure~\ref{fig:SFR_dist} we plot the SFR distribution of simulated galaxies as a function of stellar mass both in the field (all galaxies) and cluster (all galaxies belonging to the selected FOF groups using our synthetic GOGREEN halo selection described in Section \ref{sec:halo_selection_fq_desig}) environments. For convenience of plotting, any simulated galaxy which has ${\rm SFR} < 10^{-5}~\msun~{\rm yr}^{-1}$ (including those with ${\rm SFR}=0~\msun~{\rm yr}^{-1}$) is randomly assigned a logarithmic SFR value between $-4$ and $-5$ in Figure~\ref{fig:SFR_dist}.  Note that the apparent strong truncation of the distributions at low $M_{*}$ and SFR for BAHAMAS/MACSIS simulations (i.e., the rectangular edge to the distribution) is due to their lower mass resolution, which imposes a relatively large minimum (non-zero) star formation rate that can be resolved in the simulations. 

As shown in Figure~\ref{fig:SFR_dist}, all three simulations have distinct populations of star-forming galaxies (i.e., the `blue cloud', or star-forming main sequence) and a long `tail' of low-SFR galaxies, many of which have $\mathrm{SFR = 0}$. Our approach to distinguishing the star-forming and quiescent populations is straightforward, corresponding to a simple linear relation in $\log_{10}(SFR)$--$\log_{10}(M_{\rm *,30kpc})$ (or a power law in linear space) above which a galaxy is deemed to be star-forming and below which is deemed to be quiescent.  To specify this linear relation, we proceed as follows.  First, the galaxy SFRs are binned by stellar mass, and a Gaussian function is fit to the high-SFR side of the peak value to define the location (i.e., the mean of Gaussian) and the width of the star-forming main sequence (SFMS).  Note that we use the high-SFR side in order to avoid the `green valley' tail from skewing our estimate of the location of the main sequence.  With a relation for the mean SFR of the SFMS in bins of stellar mass, we simply subtract $\mathrm{3\sigma}$ from the mean SFR in each bin and fit a linear relation (in log space) to the binned data.  In other words, the amplitude of the linear relation that we use to assign star-forming status is $\mathrm{3\sigma}$ below the peak of the SFMS, while the slope of the relation matches that of the SFMS.
This division is indicated by the lines (a linear fit to the binned measurements) in the figure; with the BAHAMAS line repeated in all panels for comparison. 

We have experimented with other multiples of $\mathrm{\sigma}$ in order to gauge the impact this choice makes on the various results. In the following plots, lines correspond to the choice of $\mathrm{3\sigma}$, whereas shaded regions indicate the changes resulting from the cut being $\mathrm{1\sigma}$ and $\mathrm{5\sigma}$, both quite aggressive choices in either direction.  Note that the default choice of $\mathrm{3\sigma}$ results in a near identical match to the results of \citet{Donnari_etal_2021} for TNG300.

We note that the normalization of the SFMS is not  the same for all three models, nor is it precisely the same for the cluster and field populations within the same simulation. The latter effect, which is typically not observed, is particularly strong in EAGLE/Hydrangea.  We note that if one restricts the selection of Hydrangea galaxies to be outside the main cluster in the zoom, the location of the Hydrangea SFMS aligns very well with that of the field SFMS in EAGLE.  Thus, the difference in the position of the SFMS in/near the massive cluster is a real environmental effect in that simulation. For consistency with the analyses of the other simulations and the observations, we nevertheless use the field population to determine the boundary between quenched and star-forming galaxies for the Hydrangea cluster environment.  But we discuss below the impact of using instead the cluster SFMS to differentiate star-forming and quenched galaxies in Hydrangea.
 
\subsection{Important systematic variations in comparing simulations to observations}\label{sec-systematics}

There are at least two important systematic differences between the way the data are treated compared with the simulations, that can have some impact on our results.

The first is that the observational selection of quiescent galaxies, based on UVJ colours, is not identical to the SFR selection used in the simulations.  However, the division into two populations is largely motivated on the existence of a bimodality in observed properties, with a gap between the star-forming and quiescent galaxies, however they are defined.  We therefore expect that whether the data are classified according to colour or SFR should not make a large difference to the results
\citep[e.g.][]{2021arXiv211004314L}.  For the simulations we show on all relevant plots the uncertainty associated with varying the SFR-division within a wide range, which should account for much of the systematic uncertainty associated with this comparison.

Another difference between observations and simulations is the method used to identify cluster members: the simulations use a 3D FOF algorithm, whereas the GOGREEN cluster galaxies are selected using a circular aperture centred on the BCG and a cut in line-of-sight velocity \citep{vdBurg_etal_2020}. There is the potential for introducing a bias due to false classification of (primarily star-forming) field galaxies as cluster members, and (primarily quiescent) cluster galaxies as field. This can lead to either elevated or suppressed quenched fractions. We test this by implementing the observational selection in BAHAMAS and contrasting it with the FOF selection in Appendix \ref{app:obs_selection}.  In short, we find that such interloper contamination is minimal for a sample of clusters as massive as that in GOGREEN.   

\section{RESULTS}
\label{sec:results}

Below we compare the simulations introduced in Section \ref{sec:sims} with observations of field and cluster galaxies at $1 \lesssim z \lesssim 1.4$ described in Section \ref{sec:observations}.  We first examine the stellar mass content of central galaxies in Section \ref{sec:mstar_mhalo}, before considering the GSMF as a function of galaxy type and environment. 
Our main result is that all simulations struggle to match the observed quenched fraction in these clusters, which is demonstrated in Section \ref{sec:quench}.

\subsection{Stellar mass and SFR of central galaxies}
\label{sec:mstar_mhalo}

\begin{figure}
\includegraphics[width=\columnwidth]{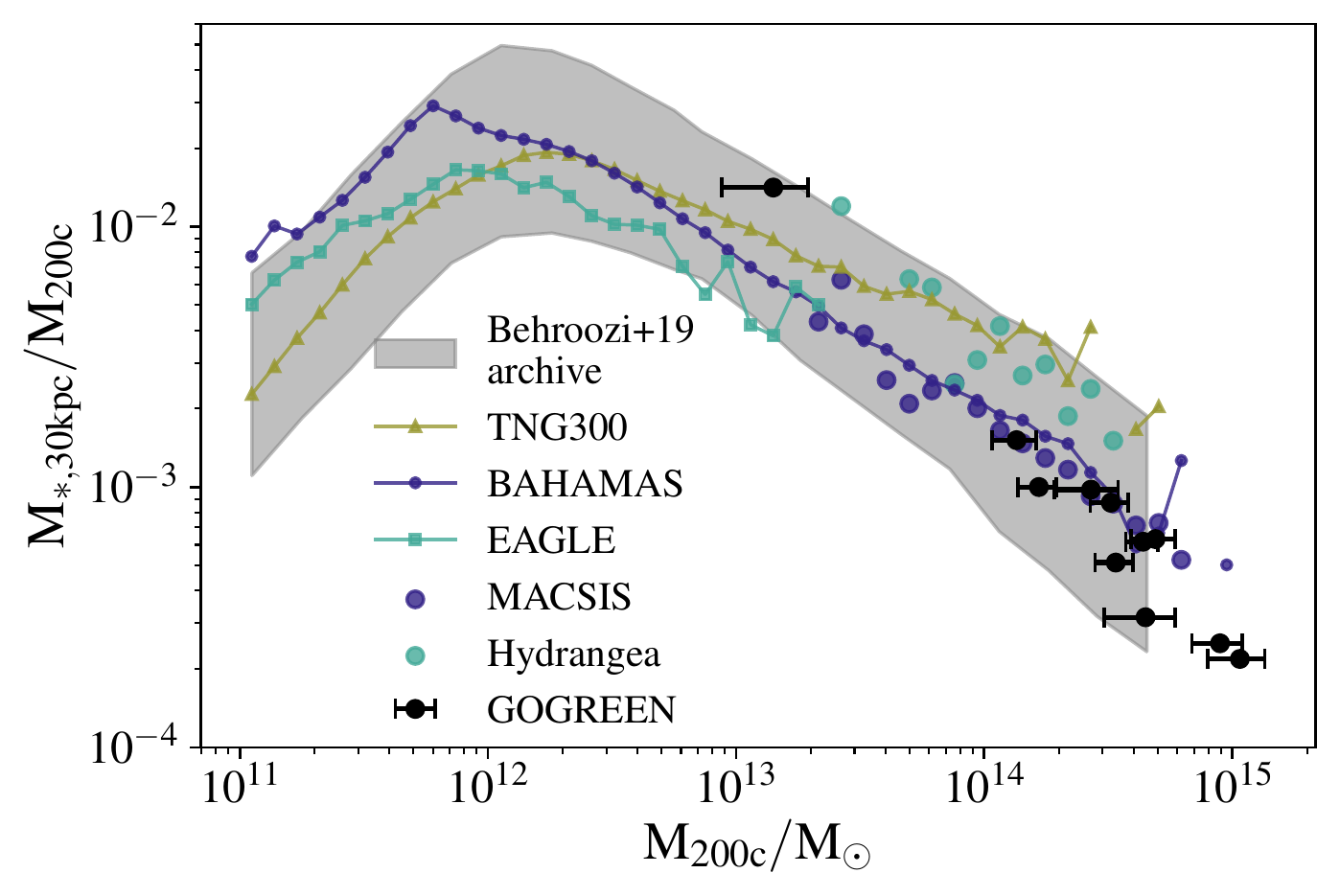}
\caption{The stellar mass fraction--halo mass relation of central galaxies at $z\approx 1$.  For the simulations, stellar masses are computed within a spherical aperture of $\mathrm{R = 30~kpc}$ (physical). Solid lines (median within the cosmological volume) and scatter points (zoom-in volumes) represent the hydrodynamical simulations: yellow for TNG300, navy for BAHAMAS/MACSIS, and turquoise for EAGLE/Hydrangea. The shaded grey region represents the full extent of `empirical model' results shown in fig. 35 of \citet{Behroozi_etal_2019}. Black points with error bars represent the 11 GOGREEN haloes.}
\label{fig:mstar_mhalo_cent}
\end{figure}

In Figure~\ref{fig:mstar_mhalo_cent} we show the stellar mass fraction--halo mass relations at $z\approx1$ for central galaxies in the simulations and make comparisons both with a sample of archival results at this epoch, as well as for the central galaxies of the GOGREEN sample.  The shaded, grey region represents the full range of constraints shown in fig.~35 of \citet{Behroozi_etal_2019} (see the caption of that figure for a full list with references).  These include abundance matching, empirical modelling, Halo Occupation Distribution modelling, and Conditional Stellar Mass Function modelling.  For this comparison, we consider the minimum and maximum values of stellar mass fraction for any model at set values of virial masses.  We then convert the default virial masses to our mass definition, $M_{200c}$, by assuming an NFW profile and adopting the mass--concentration relation of \citet{Ludlow_etal_2016}.  The conversion was done using the Colossus toolkit \citep{Diemer_2018}. The GOGREEN central galaxies are identified as the most massive galaxy with a redshift consistent with the cluster mean redshift, and projected within 500 kpc from the main galaxy over-density \citep{vdBurg_etal_2020}.  Central galaxies in the simulations and in the empirical models correspond to the stellar component of the most massive subhalo in a FOF group.  This does not necessarily have to correspond to the subhalo with the highest stellar mass, but for the vast majority of systems that is the case.   

In this specific comparison, we do not limit the analysis to massive groups/clusters as we do later, but instead we allow the comparison here to extend down to $\sim L^*$ galaxies.  By doing so we can assess whether the simulated field/central galaxies that could potentially become satellites have approximately the correct properties prior to joining the group/cluster environment. 

All curves peak at $\mathrm{log_{10}(M_{200c}/M_{\odot}) \approx 12.1}$ and decline with an approximately constant logarithmic gradient at higher halo masses, though there are small quantitative differences between the simulations. Broadly speaking, all of the simulations show reasonable agreement with the observations, to within a factor of $\sim 2$.  The systematic uncertainties that lead to variations between the observational estimates shown are comparable in magnitude to the variations between the different simulations. They also fall within the constraints defined by other types of modelling considered in \citet{Behroozi_etal_2019}.

In Appendix \ref{app:stellar_content} we examine the impact of aperture choice on the stellar mass content of centrals and on the integrated stellar masses of groups and clusters.  In short, we find that BAHAMAS/MACSIS central galaxies tend to be more spatially-extended than those of EAGLE/Hydrangea and TNG300 (which is not unexpected given the lower spatial resolution), while all simulations have similar integrated stellar masses for groups and clusters at $z \approx 1$.

\citet{McCarthy_etal_2017} have shown that the BAHAMAS simulations reproduce the observed sSFR--stellar mass relation at $z\approx1-1.5$ rather well, though the evolution of its normalization is weaker than observed (see figure~15 of that study).  \citet{Donnari_etal_2019} and \citet{Furlong_etal_2015} have also shown that the simulated SFMS (in TNG300 and EAGLE, respectively) are in reasonable agreement with the observations, though in both cases the normalization is lower by about a factor of 2 at all redshifts.

\subsection{Galaxy stellar mass functions (GSMFs)}
\label{sec:gsmf}

\begin{figure*}

\includegraphics[width=\textwidth]{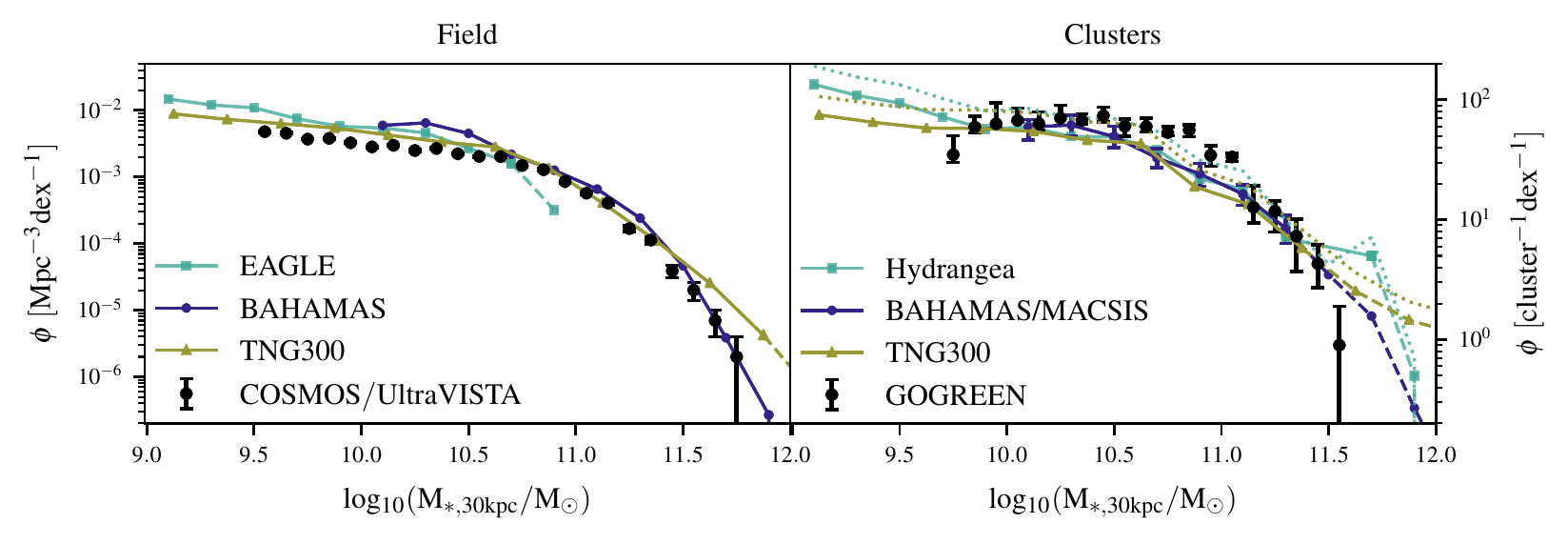}
\caption{Total stellar mass functions for the simulations (lines) and COSMOS/UltraVISTA and GOGREEN data (points with error bars) are shown for the field (left panel) and cluster (right panel) samples.  Navy error bars in the BAHAMAS/MACSIS Cluster panel represent a $1\sigma$ scatter region around the median computed from 100 GOGREEN-like 10 cluster samples. The dashed lines indicate where there are fewer than ten galaxies per bin in the simulations. The dotted lines represent the scaled versions of TNG300 and Hydrangea cluster GSMFs to account for differences in their cluster mass distributions with respect to GOGREEN.}
\label{fig:panel_gsmf}
\end{figure*}

We now examine the GSMF -- total, as well as split into quiescent and star-forming populations -- and its variations between the field and cluster environments.

\subsubsection{The total stellar mass function}

Figure~\ref{fig:panel_gsmf} shows the measured GSMFs for all three simulations, compared with those from COSMOS/UltraVISTA and GOGREEN as measured by \citet{Muzzin_etal_2013, vdBurg_etal_2020}. Field measurements are in the left panel, and cluster GSMFs are in the right panel.  The cluster GSMF represents the number of galaxies per stellar mass bin (dex$^{-1}$) per cluster, obtained by stacking (summing) the individual GSMFs of each cluster and dividing through by the number of clusters in the stack.  For BAHAMAS/MACSIS, 100 samples of 10 haloes are drawn (matching the GOGREEN distribution) and GSMFs estimated. We plot the median value in bins of stellar mass with 1$\sigma$ scatter region represented by navy error bars.  The Hydrangea and TNG300 simulation samples described above are represented by solid lines, switching to dashed lines when there are fewer than ten galaxies in a mass bin. The GOGREEN observations are represented by black data points with error bars.

All the models were calibrated in part to reproduce the field GSMF (at least at lower redshifts), and this is reflected in the good agreement with the data shown in the left panel of Figure~\ref{fig:panel_gsmf}.  
On the other hand, the shape of the GOGREEN cluster GSMF (right panel) is not particularly well reproduced in any of the simulations.  A more detailed discussion of this is given in Appendix~\ref{app:gsmfs}; here we briefly discuss the main differences.   One notable discrepancy is near the knee ($\mathrm{10.5 < \log_{10}(M_*/M_{\odot}) < 11.2}$), where all simulations underpredict the observed number of galaxies.  We suspect that in BAHAMAS this may be related to the lower resolution.  However, in TNG300 and Hydrangea this is largely explained as a small mismatch in the halo mass distribution compared with the observations.  Using the BAHAMAS/MACSIS suite we have derived an approximate scaling factor to scale the GSMF from the Hydrangea and TNG samples to that of a sample with the GOGREEN mean halo mass.  We show the scaled GSMFs in Figure~\ref{fig:panel_gsmf} with the dotted cyan curve, which is in much better agreement with GOGREEN near the knee, and consistent with what \citet{Ahad2021} found for Hydrangea. Scaling the curves up by this factor does, however, exacerbate the differences with respect to GOGREEN at the very lowest and highest masses. In other words, while Hydrangea and TNG300 reproduce the amplitude of the cluster GSMF relatively well (once the difference in halo mass is accounted for), the shape of the predicted GSMF differs in detail from that observed in GOGREEN clusters.

\begin{figure*}
\includegraphics[width=\textwidth]{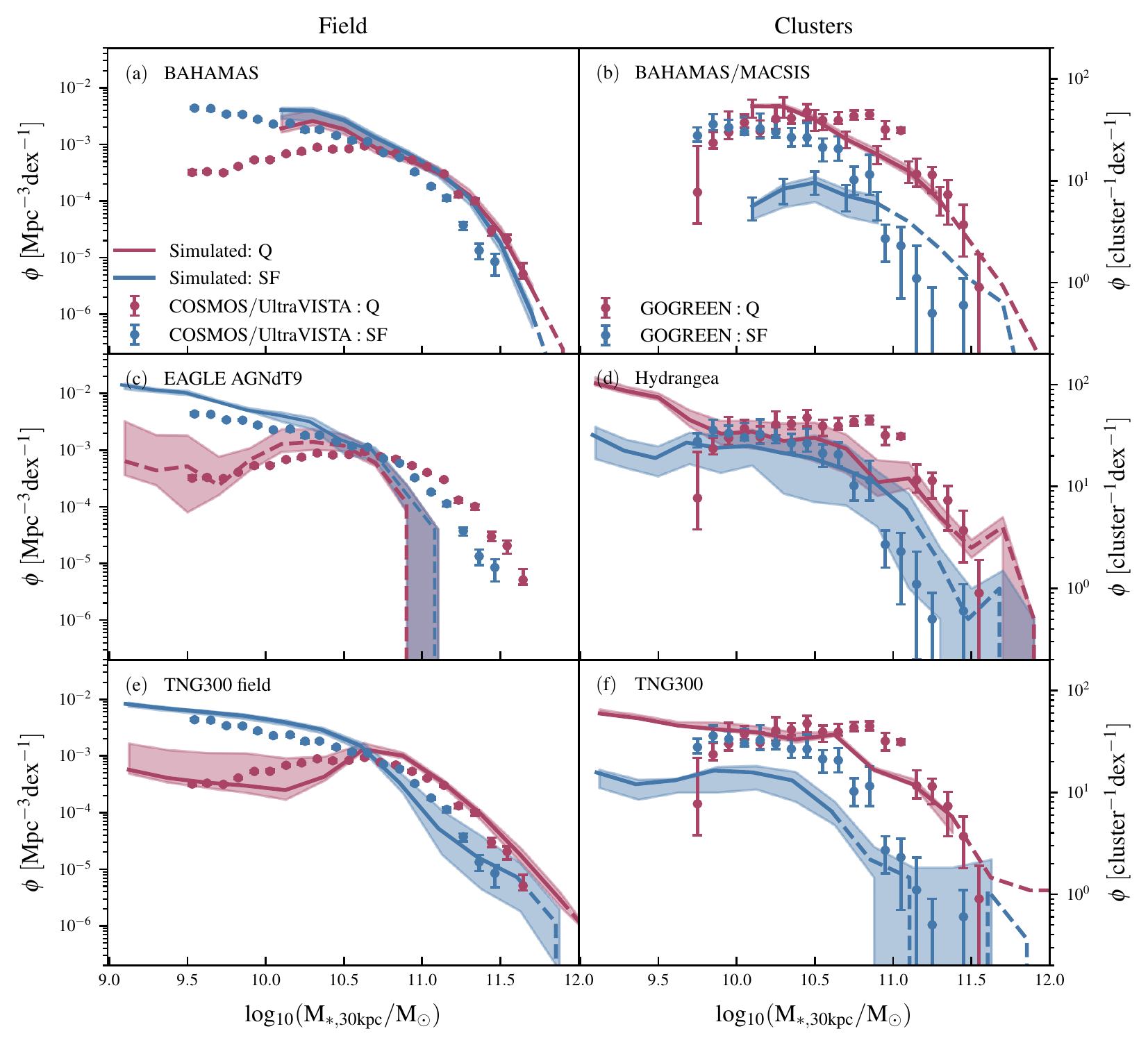}
\caption{Field (left column) and cluster (right column) GSMFs are shown for the three simulation models with lines.  The lines are dashed in regions with fewer than ten galaxies.  Each row shows a different simulation, as indicated.  The observations are represented by points with error bars, and are the same for all rows in the respective Field/Cluster columns. Blue lines/points represent star-forming galaxies, while red indicates quenched galaxies.   Shaded regions indicate the possible variation as a result of star-forming--quenched division choice (see text).  Blue/red error bars in the BAHAMAS/MACSIS panel represent a $1\sigma$ scatter region around the median computed from 100 GOGREEN-like 10 cluster samples. 
}
\label{fig:panel_gsmf_2}
\end{figure*}

\subsubsection{Stellar mass functions for star-forming and quiescent galaxies}

In Figure~\ref{fig:panel_gsmf_2} we show the GSMFs in the data and simulations, now split according to star-forming vs.~quiescent status.  As previously shown in \citet{vdBurg_etal_2020}, the observed GSMFs of star-forming and quiescent galaxies in the field have distinct shapes, and in fact cross at $\mathrm{log_{10}(M_*/M_{\odot}) \approx 10.75}$, with star-forming galaxies dominating the low-mass end and quiescent galaxies dominating at the highest masses. In the GOGREEN clusters, the shapes of both the star-forming and quenched GSMFs are the same in the cluster as they are in the field; only the relative normalization of the quenched GSMF is much higher in the cluster environment.   The high-mass end is completely dominated by quiescent galaxies, whereas at the low-mass end the two have comparable amplitudes down to the lowest measured stellar masses ($\mathrm{log_{10}(M_*/M_{\odot}) \approx 9.5}$). 

In general, all three models do a reasonable job of replicating the qualitative trends in the field, in the sense that quiescent galaxies are more abundant at the high mass end.  BAHAMAS overpredicts the abundance of massive star-forming galaxies and of quiescent, low mass galaxies. EAGLE reproduces the observed field trends reasonably well, at least in the stellar mass range $\mathrm{log_{10}(M_*/M_{\odot}) \lesssim 10.75}$. It overestimates the star-forming galaxy abundances by a factor of a few but matches the slopes and cross-over point of the two curves.  Finally, TNG300 shows, qualitatively, the best match to the observations in the field, matching the gradients, amplitudes, and cross-over point between star-forming and quiescent populations; this may be expected given that aspects of the feedback were adjusted to better reproduce the evolution of the luminosity functions in different passbands. The match is not exact, being off by a factor of a few in some places, but much of it can be accounted for by the uncertainty in selecting the division between star-forming and quiescent galaxies (indicated by the shaded region). It is interesting to note that TNG300 has a significant intermediate galaxy population, suggesting that quenching is a more gradual process relative to the other two simulations. This makes it more sensitive to the definition of quiescence described in Section \ref{sec:sims-sfr}. 

The picture is quite different in the clusters, where both BAHAMAS and TNG300 predict a much lower abundance of star-forming galaxies than observed.  Moreover, the GSMF for cluster star-forming galaxies in BAHAMAS is much flatter than the observed one.  The Hydrangea simulations generally provide a better match to the data, though the abundance of quiescent galaxies near the knee of the mass functions is underestimated by a factor of a few.  It is also notable that Hydrangea and TNG300 differ in the predicted behaviour for masses below the mass limit of the data, with Hydrangea predicting a steeper slope for the quiescent galaxy mass function than TNG300.

The differences between simulations are more clearly seen when we consider quenched fractions, in the following section.  We therefore defer further detailed discussion of this figure to Appendix~\ref{app:gsmfs}.

\begin{figure*}
\includegraphics[width=\textwidth]{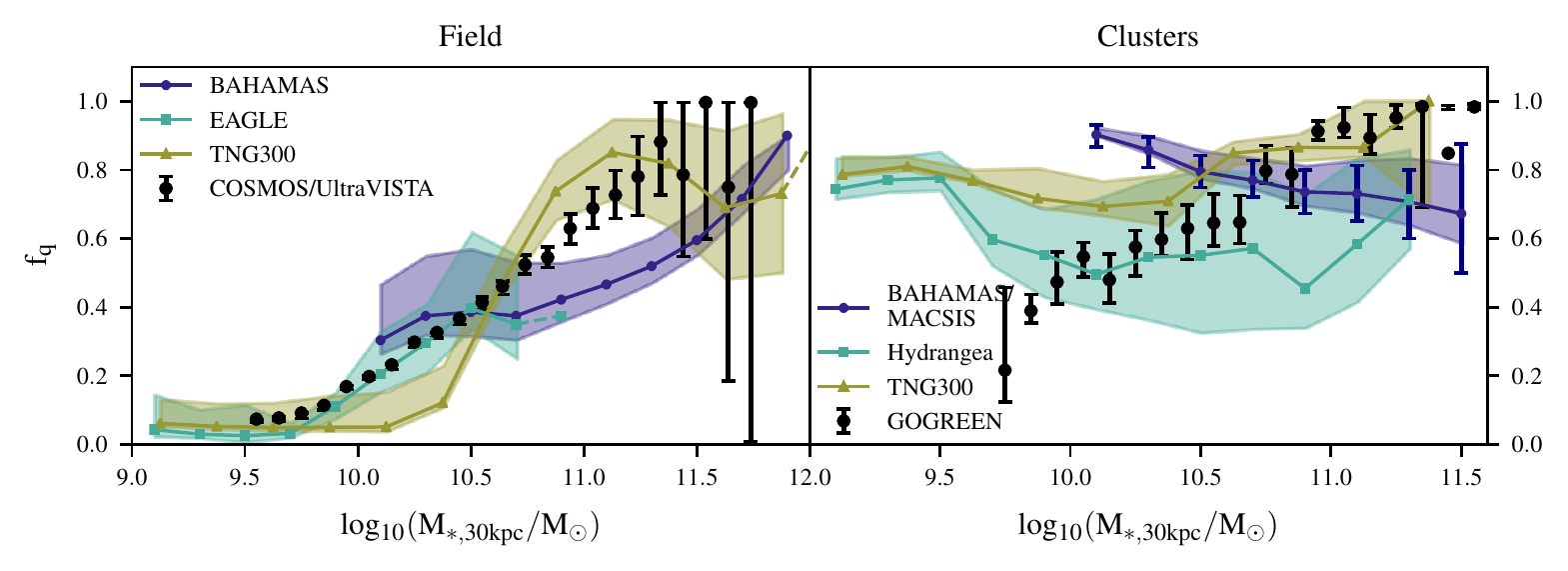}
\caption{The fraction of quenched galaxies, $\mathrm{f_q}$ is shown as a function of stellar mass for the three simulations (solid/dashed lines) and the COSMOS/UltraVista (Field, left panel) and GOGREEN (cluster, right panel) observations as indicated in the legends. Transparent data points in the right panel show the scaled version of GOGREEN where selection effects have been accounted for (see Appendix~\ref{app:obs_selection}). Navy error bars in the BAHAMAS/MACSIS Cluster panel represent a $1\sigma$ scatter region around the median computed from 100 GOGREEN-like 10 cluster samples. Shaded regions indicate the maximum variation expected from the choice of star-forming--quenched division.  Both the normalization and the trend with mass in clusters is in poor agreement with the data.}
\label{fig:qf}
\end{figure*}

\subsection{The fraction of quenched galaxies}
\label{sec:quench}

The quenched fraction, $\mathrm{f_q}$, is defined simply as the ratio of the number of quiescent galaxies to the total number of galaxies in a given stellar mass bin\footnote{We note that the rescaling applied to the Hydrangea and TNG300 GSMFs in Section~\ref{sec:gsmf} to better match the mean halo mass of GOGREEN is not carried through to the measurement of the quenched fraction here. We make the assumption that star-forming and quiescent GSMFs are affected equally and, with quenched fraction being a relative quantity, the global renormalization factors out. In reality the quenched fraction is expected to be halo mass-dependent \citep{Weinmann_etal_2006, Wetzel_etal_2012}, so our Hydrangea and TNG300 cluster quenched fractions may be slightly underestimated compared to a case where their mean halo masses were slightly higher and a better match to the GOGREEN sample.   However, the differences would be small relative to the trends derived below and our conclusions are conservative, in that we find that the simulations are already too efficient at quenching satellites in clusters.} and is shown in Figure~\ref{fig:qf}.   

Focusing first on the field population, shown in the left panel, all three simulations predict an increase of the quenched fractions with galaxy stellar mass, in qualitative agreement with data. We note that this measurement is sensitive to the quantitative distinction between star-forming and quiescent: the choice of how far the cut lies from the SFMS (number of $\sigma$) can change the $\mathrm{f_q}$ estimate by $\mathrm{\pm 0.1}$ in all three simulations, as indicated by the shaded regions. This is in addition to the uncertainty associated with determining the designation in the first place (discussed in Section~\ref{sec:halo_selection_fq_desig}).
BAHAMAS reproduces the general trend, however, it underpredicts the field quenched fraction over most of the stellar mass range by $\approx0.2$. As it is generally thought that AGN feedback is responsible for the quenching of massive galaxies in nature, the trends in left panel of Figure~\ref{fig:qf} may suggest that AGN feedback is not efficient enough, specifically with regards to halting star formation at these redshifts. It performs better at the lowest stellar masses. EAGLE is the best-performing simulation on this metric: it matches the observed trend very well over the range where the number of galaxies is sufficiently high to sample. It has been shown to reproduce this trend at $z=0$ \citep{Furlong_etal_2015} and we see no indication that this does not hold at $z=1$.  TNG300 reproduces the overall increasing trend but the gradient varies substantially as a function of stellar mass, likely owing to the different modes of AGN feedback dominating at certain times.

In the clusters, none of the simulations reproduce the observed correlation between $\mathrm{f_q}$ and stellar mass in detail. Only TNG300 comes close to reproducing the upward trend, and that only comes into effect at $\mathrm{log_{10}(M_*/M_{\odot}) \gtrsim 10.4}$. Below this stellar mass, $\mathrm{f_q}$ stays at a high value of $\sim0.8$ down to the lowest stellar masses. BAHAMAS/MACSIS and Hydrangea exhibit opposite trends with stellar mass to what is observed: $\mathrm{f_q}$ is higher at lower stellar masses than it is at the highest. Low-mass satellites are clearly being quenched too easily, while a high fraction of centrals are star-forming instead of being quenched.

We remind the reader that quiescent galaxies are defined differently in the simulations (based on SFR) and observations (UVJ colour).  While this could quantitatively affect the normalization of $\mathrm{f_q}$, it is not likely to have a strong effect on the trends with stellar mass.  Moreover, the fact that the $\mathrm{f_q}$ in the simulated field population matches the observations fairly well makes it appear unlikely that the large difference observed in clusters can be attributed to this difference in definition. This same over-quenching of satellite galaxies in EAGLE and BAHAMAS has been identified in \citet{Kukstas_etal_2020} via different means, identifying the hot gas properties as the primary cause. That study considered galaxies at redshifts up to $z=0.15$, and here we show that the same issue exists at $z \approx 1.0$.

As also noted previously, the observational selection of cluster members differs from the FOF selection in the simulations (Section \ref{sec-systematics}).  This both excludes cluster members and includes a potentially substantial number of field galaxies from the data (relative to the simulation definition) that could lead to a bias in the observationally inferred quenched fractions.  However, as we quantitatively show in Appendix~\ref{app:obs_selection} using both simple analytic calculations and the BAHAMAS simulations, the bias in the recovered quenched fraction is expected to be very small in comparison to the other uncertainties we have already discussed (e.g., location of the SFMS and number of $\sigma$ used for the cut).  This is primarily due to the fact that GOGREEN consists of very massive systems whose abundant satellite populations greatly exceed the number of interlopers along the line of sight.  Furthermore, the 1 Mpc radius aperture is well suited to the masses of the systems under consideration, so few genuine cluster members are excluded by this selection criterion.

\subsubsection{Quenched fraction excess}

\begin{figure}
\includegraphics[width=\columnwidth]{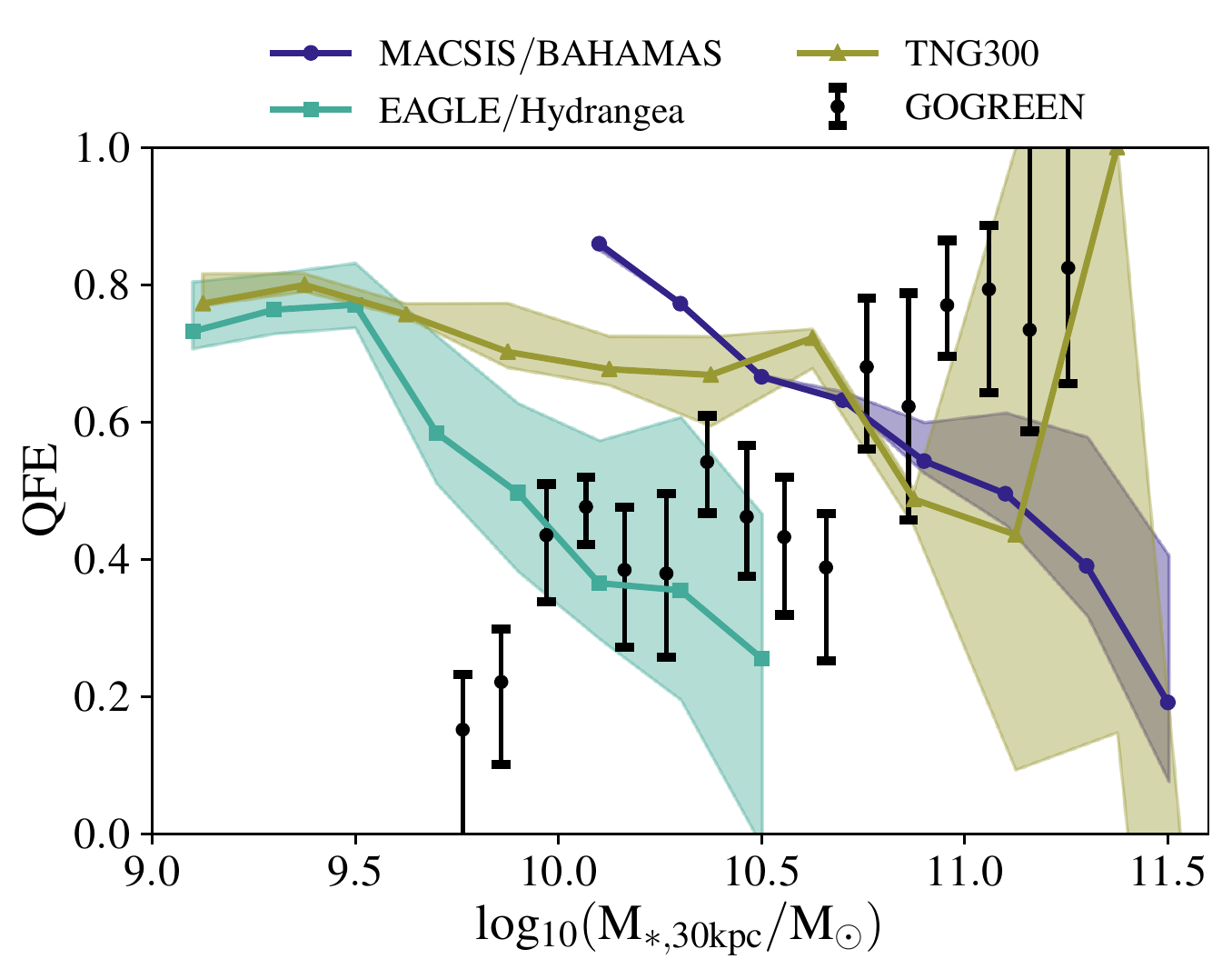}
\caption{The quenched fraction excess, QFE, is shown as a function of stellar mass for the models and data as in Figure~\ref{fig:qf}.  This quantity aims to measure the amount of quenching in clusters relative to the field, and is thus useful for isolating the effect of clusters.  Transparent points have been corrected for systematic contamination as discussed in Appendix \ref{app:obs_selection}; they have been shifted to the right for visibility. Shaded regions indicate the maximum variation expected from the choice of star-forming--quenched division.}
\label{fig:qfe}
\end{figure}

One way to try and isolate the quenching physics associated with clusters is to compute the `quenched fraction excess' (QFE): $\mathrm{(f_q^{cluster} - f_q^{field})/(1-f_q^{field})}$, as proposed by \citet{van_den_Bosch_2008}, \citet{Wetzel_etal_2012}, and others.  By normalizing relative to the field, this quantity highlights differences in $\mathrm{f_q}$ that are correlated with environment, though its interpretation in detail is non-trivial \citep[see, for example, Appendix~A in ][]{McNab}.
We show this quantity in Figure~\ref{fig:qfe} as a function of stellar mass.  Contrary to what is observed at low redshift \citep{Wetzel_etal_2012}, the $z>1$ GOGREEN data show a strongly increasing QFE with satellite mass \citep{vdBurg_etal_2020}. 

Again, all three simulations fail to match the data.  The same feature of declining gradient that was seen in cluster $\mathrm{f_q}$ carries through.  The QFE in BAHAMAS is strongly declining with stellar mass. This is despite the fact that $\mathrm{f_q^{field}}$ mirrors the shape of observations with $\mathrm{log_{10}(M_*)}$. The anti-correlation seen in $\mathrm{f_q^{cluster}}$ is emphasised here. 
A similar result is observed in the EAGLE/Hydrangea simulations: we see the same anti-correlation as for BAHAMAS, albeit intersecting the GOGREEN data at a different stellar mass value. This decline is driven exclusively by $\mathrm{f_q^{cluster}}$ and is in line with what has been reported by \citet{Bahe_etal_2017_hydrangea} for $z\approx0$: the QFE is excessively high at low stellar masses but declines to match observations at higher $\mathrm{log_{10}(M_*)}$. The only difference we see here is that the decline starts at lower stellar masses.  TNG300 samples the entire $\mathrm{9.0 < log_{10}(M_*) < 11.5}$ stellar mass range and provides the best match to the observations at stellar masses $\mathrm{log_{10}(M_*/M_{\odot}) > 10.0}$. It still suffers from the same over-quenching problem at low stellar masses and exhibits an overall declining trend.

\section{Discussion and Conclusions}
\label{sec:conclusions}

The subgrid physics parameters governing the efficiencies of feedback processes in most modern cosmological hydrodynamical simulations are calibrated on observed properties of the global galaxy population.   In that sense, comparisons with galaxies in rare, dense environments provide a useful and necessary test of those models.  Most such comparisons to date have been limited to low redshifts $z<1$, where there is a wealth of data.  While successful in some regards, those first comparisons also showed some interesting discrepancies related to over-quenching of low-mass satellites (e.g., \citealt{Vulcani14,Bahe_etal_2017,Kukstas_etal_2020}).

There is now growing evidence that environmental quenching mechanisms may be different at $z\sim 1$ from those observed in the local Universe.  The empirical correlations between the fraction of star-forming galaxies, their stellar masses, and their host environments are very different from what they are at $z=0$ \citep[e.g.][]{Muzzin_etal_2014,Balogh_etal_2016,Kawinwanichakij_etal_2017}.  This provides a new opportunity for semi-independent tests of the models, where disagreements with the data may inform and guide further improvements in the underlying physics.  To this end, we have taken advantage of the large, high quality, spectroscopic observations of $1<z<1.4$ clusters from GOGREEN to test three different implementations of physical models: BAHAMAS, EAGLE, and TNG300 as periodic boxes and MACSIS, Hydrangea as zoom-in simulations.  The MACSIS and Hydrangea zoom suites were designed to complement BAHAMAS and EAGLE, respectively.

Specifically, we compare the GSMFs of star-forming and quiescent galaxies in the cluster and field environment at $1<z<1.4$, and make the following key observations:
\begin{itemize}
\item All three models reproduce the field GSMFs qualitatively well, including when separated into star-forming and quiescent galaxies.  Necessarily, then, they also qualitatively reproduce the observed correlation between the fraction of quiescent galaxies $\mathrm{f_q}$ and stellar mass.   However, BAHAMAS predicts too many high-mass ($\mathrm{log_{10}(M_*/M_{\odot}) > 11.0}$ ) star-forming galaxies, leading to a quenched fraction that increases more slowly with stellar mass than the data, and is thus too low at high masses.  
\item All predict similar total GSMFs in $z=1$ clusters.  While they agree tolerably well with the data, they do not show a strong break and thus predict both too many massive galaxies (many of which are centrals), and too many low mass galaxies, relative to the abundance at $M^\ast$.  
\item All models predict a steep low-mass slope to the quiescent galaxy GSMF in clusters, and this population dominates at low stellar masses (though the precise mass scale varies significantly between the simulations).  This is not observed in the GOGREEN data, suggesting that there is still a significant over-quenching problem in the simulations at $z\approx1$. 
\item The dependence of quenched fraction, $\mathrm{f_q}$, on stellar mass in clusters is very different in all three simulations, and none provide a good overall match to the data.  TNG300 provides good agreement for $\mathrm{log_{10}(M_*/M_{\odot}) > 10.5}$, but predicts $\mathrm{f_q}$ should increase toward low masses, while the observed $\mathrm{f_q}$ decreases.   Both BAHAMAS/MACSIS and EAGLE/Hydrangea exhibit a correlation between $\mathrm{f_q}$ and stellar mass that is opposite to that observed, and with very different normalizations.  At all masses  $\mathrm{log_{10}(M_*/M_{\odot}) > 10.5}$, the variation in $\mathrm{f_q}$ between the three simulations ranges from $\approx 0.2$ to $>0.8$.
\item None of the simulations reproduce the observed positive correlation between the quenched fraction excess (QFE) and stellar mass.
\end{itemize}

The mismatch between the observed cluster $f_q$ in the data and simulations, particularly at relatively low stellar masses, presents a clear opportunity to identify missing or mischaracterized physics in the simulations.  
The fact that the three models differ from one another in detail means we can look for differences in their nature for clues.  

One possibility is that limited numerical resolution and the lack of an explicit modelling of the cold ISM in each of these simulations results in overly efficient quenching with respect to real low-mass satellite galaxies.  In this regard, it is noteworthy that the stellar mass scale where cluster over-quenching kicks in is ordered by resolution (BAHAMAS/MACSIS, TNG300, EAGLE/Hydrangea).  Note that finite resolution and the lack of a cold ISM may impact environmental quenching in several connected ways.  First, it is clear that the gravitational potential wells of low-mass galaxies are relatively less well resolved, which generally means that the inner mass distribution will be too extended and therefore artificially susceptible to tidal forces. Idealised simulations by \citet{van_den_Bosch_Ogiya_2018} suggest that this can cause artificial tidal disruption even for massive subhaloes, although full cosmological simulations (Hydrangea) found satellite disruption (physical or artificial) in massive clusters to be restricted to the earliest-accreted galaxies \citep{Bahe_etal_2019}. Finite resolution can also result in feedback processes being more bursty and energetic, depending on the details of the implementation.  For example, the AGN feedback implementations in EAGLE/Hydrangea and BAHAMAS/MACSIS are similar, in that they heat a similar number of particles by a similar amount (i.e., similar $\Delta T_{\rm heat}$).  However, the mass resolution differs between the simulations by almost a factor of a thousand, implying the energy per feedback episode in BAHAMAS is significantly larger than that in EAGLE.  Both sets of simulations reproduce the present-day BH scaling relations relatively well (through calibration), implying that the total injected energies (integrated over cosmic time) are similar, but also implying the injection in EAGLE is much more continuous than that in BAHAMAS as a consequence of heating a fixed number of particles rather than a fixed Lagrangian region (mass).  The net result of this is that, even though the stellar masses (and to an extent the integrated gas masses) are calibrated to be similar, the radial distribution of gas in haloes could be quite different (indeed, see the comparison in \citealt{Oppenheimer_et_al_2021}), resulting in different environment quenching.  Finally, and perhaps most obvious, finite resolution and the lack of a cold ISM may result in overly efficient ram pressure stripping of low-mass simulated galaxies, as idealised simulations have shown the cold molecular phase to be significantly more resistant to ram pressure (e.g.~\citealt{Tonnesen_Bryan_2009}).

If finite resolution is indeed responsible for the tension at low masses, it implies that \textit{none} of the current simulations have sufficient resolution to cover the full mass range accessible to observations, as they all show deviations from the data at low mass.  However, without simulations of significantly higher resolution and an explicit cold ISM model, we are unable to test this hypothesis.  Given this is the case, we therefore also cannot rule out the possibility that all of the simulations are missing important physics (e.g., magnetic draping) which may help real satellites to retain their star-forming gas for a longer period of time post-infall.

One way to make further progress on the simulation side is to carry out a dedicated and systematic exploration of the effects of variations in subgrid physics and resolution on the predicted environmental trends.  Such a study would be useful not only for identifying more realistic models but also in helping to elucidate the complex relationship between feedback and environmental processing.

\section*{Data Availability}
All observational data used in this paper are available from the GOGREEN and GCLASS public data release, at 
the CADC (\url{https://www.cadc-ccda.hia-iha. nrc-cnrc.gc.ca/en/community/gogreen}), and NSF’s NOIR-Lab (\url{https://datalab.noao.edu/gogreendr1/}).
EAGLE and TNG300 simulation public data releases can be accessed on the EAGLE project website (\url{http://icc.dur.ac.uk/Eagle/database.php}) and Illustris TNG website (\url{https://www.tng-project.org/}). 

\section*{Acknowledgements}
We thank the native Hawaiians for the use of Maunakea, as observations from Gemini, CFHT, and Subaru were all used as part of our survey.  The authors thank David Barnes and Scott Kay for sharing their MACSIS simulation data with us.  MB gratefully acknowledges support from the NSERC Discovery Grant program.  This project has received funding from the European Research Council (ERC) under the European Union's Horizon 2020 research and innovation programme (grant agreement No 769130). YMB gratefully acknowledges funding from the Netherlands Organization for Scientific Research (NWO) through Veni grant number 639.041.751. GW gratefully acknowledges support from the National Science Foundation through grant AST-1517863, and from from HST program numbers GO-15294 and GO-16300. Support for program numbers GO-15294 and GO-16300 was provided by NASA through a grant from the Space Telescope Science Institute, which is operated by the Association of Universities for Research in Astronomy, Incorporated, under NASA contract NAS5-26555. RD gratefully acknowledges support by the ANID BASAL projects ACE210002 and FB210003. GR gratefully acknowledges support from NSF AST-1517815, HST program numbers GO-15294 and AR-14310, and NASA ADAP award 80NSSC19K0592. MCC acknowledges support from NSF grants AST-1518257 and AST-1815475. This work used the DiRAC@Durham facility managed by the Institute for Computational Cosmology on behalf of the STFC DiRAC HPC Facility. The equipment was funded by BEIS capital funding via STFC capital grants ST/P002293/1, ST/R002371/1 and ST/S002502/1, Durham University and STFC operations grant ST/R000832/1. DiRAC is part of the National e-Infrastructure. IPC acknowledge the financial support from the Spanish Ministry of Science and Innovation and the European Union - NextGenerationEU through the Recovery and Resilience Facility project ICTS-MRR-2021-03-CEFCA. DCB thanks the LSSTC Data Science Fellowship Program, which is funded by LSSTC, NSF Cybertraining Grant $\#$1829740, the Brinson Foundation, and the Moore Foundation; participation in the program has greatly benefited this work. FS acknowledges support by a CNES fellowship.




\bibliographystyle{mnras.bst}
\bibliography{Bibliography} 



\appendix

\section{Investigating the impact of observational selection effects}
\label{app:obs_selection}

\begin{figure*}%
    \centering
    \includegraphics[width=\textwidth]{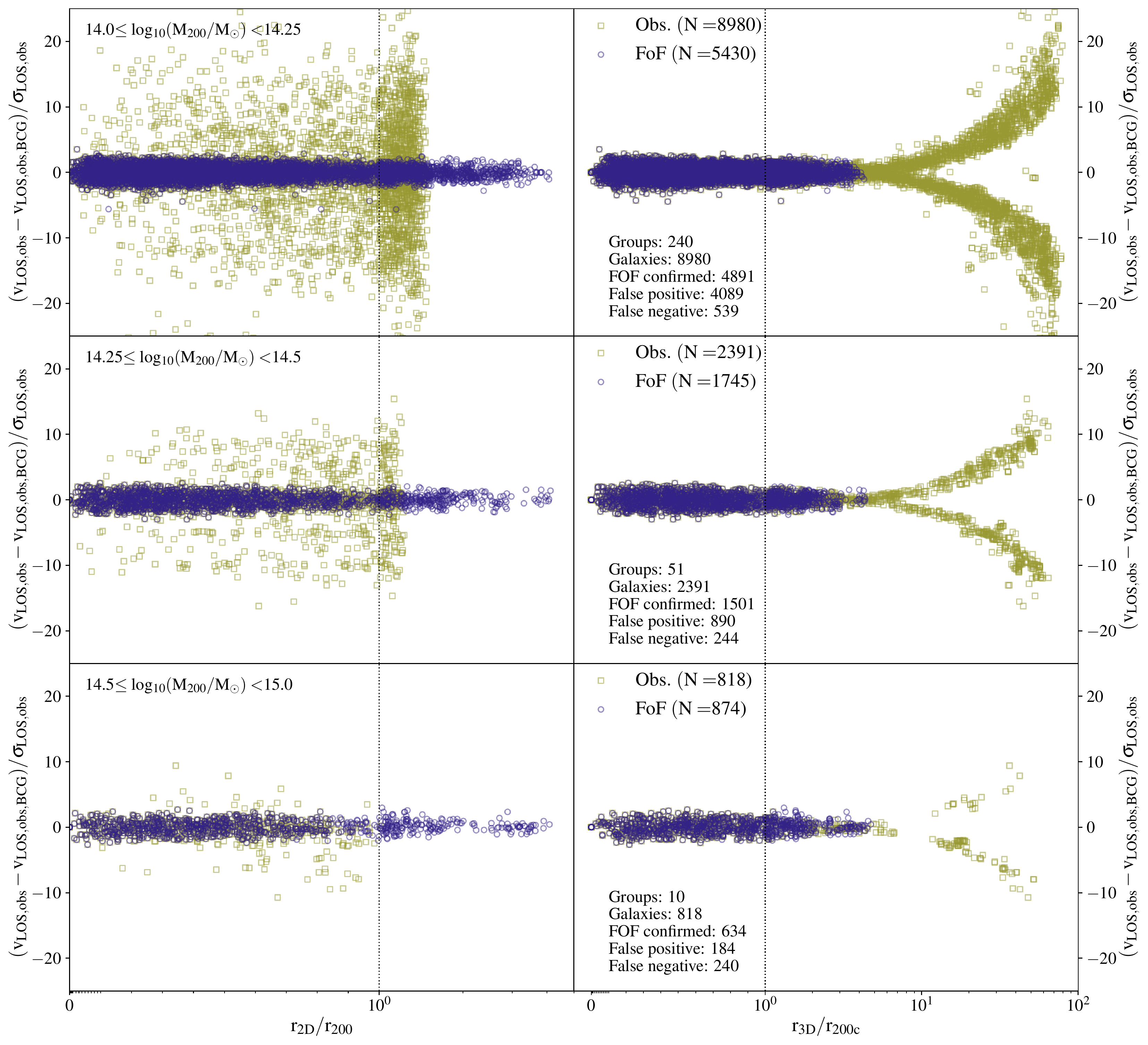}%
    \caption[Phase-space diagrams for two different galaxy membership selections in bins of halo mass.]{Phase-space diagrams for three galaxy selections in bins of halo mass. Navy circles show the FOF selection and yellow squares show the full `observational' selection of vdB20. Left column shows radial distances in two transverse dimensions; right columns show the three-dimensional radius, normalised by $\mathrm{r_{200c}}$. Rows display samples in three different halo mass bins. Note that x-axis is semi-logarithmic; it is linear in the inner regions of the cluster (within $\mathrm{r_{200c}}$) and logarithmic outside. Numbers displayed in the legends show the total number of galaxies under each selection, whereas numbers near the bottom of each panel show diagnostic information between the two methods.}%
    \label{fig:phase_space}%
\end{figure*}

\begin{figure*}%
    \includegraphics[width=\textwidth]{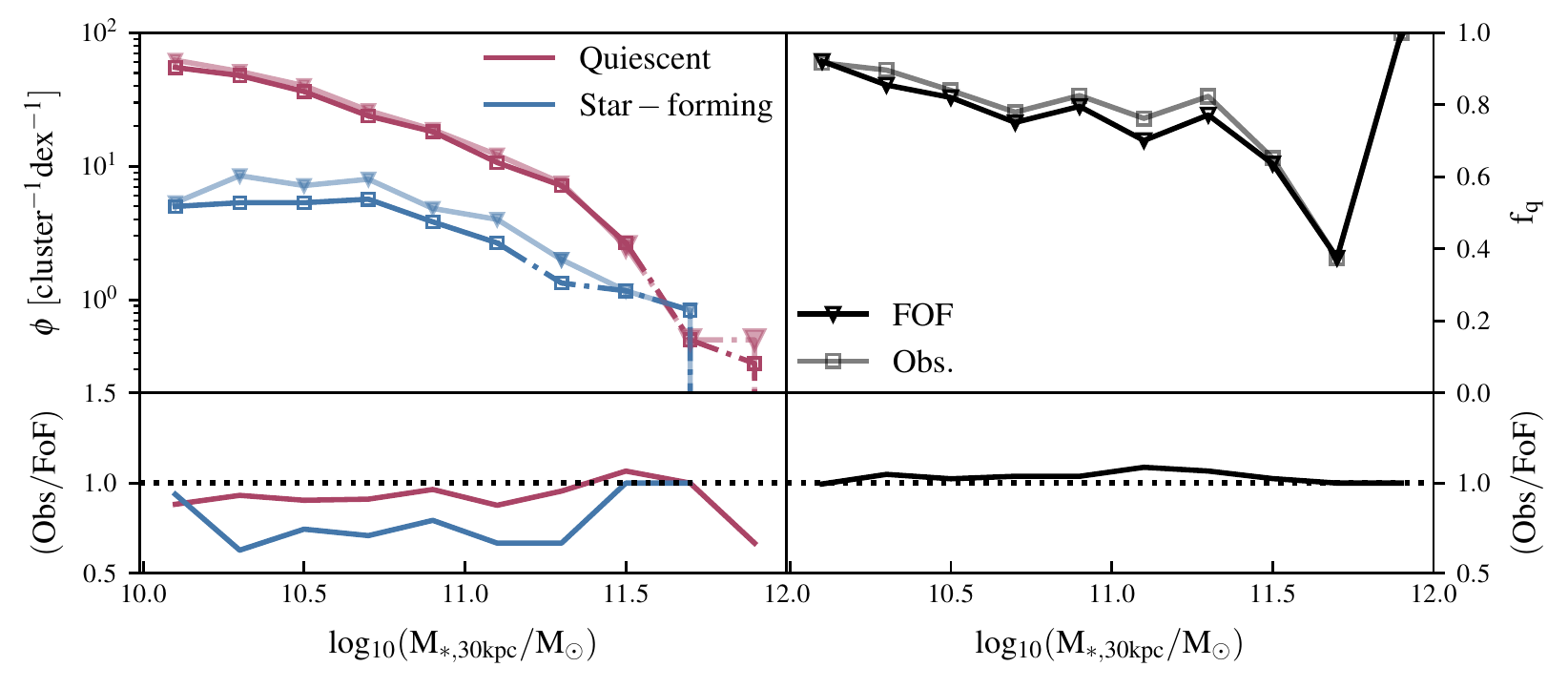}%
    \caption{Cluster GSMF measurements for observational and FOF selections, split by galaxy type (left) and quenched fraction as a function of stellar mass estimates for both selections (right). Halo mass function matches that of GOGREEN as in the main text. Lines with square markers represent the `observational' selection and lines with triangles show FOF selection. For GSMFs, blue lines represent star-forming galaxies and red lines show their quiescent counterparts. Each panel is accompanied by a ratio of observational to FOF selection for each quantity (bottom row). Observational selection excludes $\approx 25\%$ of galaxies due to a fixed aperture, but does so equally for start-forming and quiescent galaxies - leaving quenched fraction unaffected.}%
    \label{fig:gsmf_diff}%
\end{figure*}

In \citet{vdBurg_etal_2020} (hereafter vdB20), spectroscopically observed galaxies were assigned membership by introducing a circular aperture of $\mathrm{R = 1~Mpc}$, centred on the cluster BCG, and a cut in velocity relative to the cluster corresponding to a redshift difference of $\mathrm{|\Delta z| = 0.02(1+z)}$. This choice corresponds to $\approx 2-3\sigma_{LOS}$ of the most massive GOGREEN clusters. Galaxies with only photo-z estimates (non-targets, because they were not targeted for spectroscopic observation) were given a more generous $\mathrm{|\Delta z| = 0.08(1+z)}$. By assuming that spectroscopic galaxies are a representative sub-sample of the entire sample, vdB20 were able to introduce correction factors in order to correct these photometric memberships to correspond to the narrower, spectroscopic definition statistically (see Sec. 3.5 of vdB20 for a more detailed description). The effect of such an exercise is that effective membership selection applied to \emph{all} galaxies is that of spectroscopically-targeted ones, i.e. $\mathrm{R = 1~Mpc}$ and $\mathrm{|\Delta z| = 0.02(1+z)}$.

Here we address two possible sources of systematic uncertainty with this membership definition.  The first is that the spectroscopic membership definition is quite broad, corresponding to a length scale of $\pm95$ cMpc along the line of sight.  This will lead to a population of projected field galaxies, physically unassociated with the cluster, that are included in the spectroscopic membership. The second is that of the physical $\mathrm{R = 1~Mpc}$, which can exclude member galaxies by imposing a transverse distance limit. 
Using the field galaxy SMFs published in vdB20 we can estimate the magnitude of this contamination as a function of stellar mass.  The average volume of each GOGREEN cluster, taken within a 1 Mpc physical radius and $\mathrm{|\Delta z| = 0.02(1+z)}$, is 3130 cMpc$^3$.  At $\log(M_*)=10.0$ (for example) there are only $\approx 10$ star-forming galaxies, and $\approx 1$ quiescent galaxy, expected in a random field sample of this volume at $z\approx 1.2$.  The resulting correction amounts to only $\sim 5$ per cent for quiescent galaxies, and $\sim 25$ per cent of the star-forming population, with little dependence on stellar mass.  The impact on the resulting quenched fraction is correspondingly small (reducing it from $0.61$ to $0.55$ at $\log(M*)=10.0$) and does not impact any of the conclusions reached in this paper.  This calculation does neglect any correlation of large scale structure (or the galaxy populations within them), which will tend to increase the field contamination in the vicinity of the cluster (i.e., galaxy groups are more likely to be clustered near a massive cluster and groups will have a higher quenched population relative to the field).  However, the importance of this effect can be tested by applying the GOGREEN selection criteria to the simulations (as we do below), at least for the case of BAHAMAS where the volume is sufficiently large to capture all of the selected galaxies along the line of sight and for which sufficiently massive haloes exist to closely match the GOGREEN cluster selection.  

There are two main components contributing to an observed recession velocity of a distant galaxy: (i) Hubble flow associated with the expansion of the Universe and (ii) galaxy's peculiar velocity along the line of sight.
A table of comoving distances for given redshifts can be computed as (e.g., \citealt{Hogg_1999}):

\begin{equation}
    d_{c}(z) = d_{H} \int_0^z \frac{dz'}{E(z')},
\end{equation}
where $\mathrm{d_{C}(z)}$ is the comoving distance at redshift z, $\mathrm{d_{H}=c/H_{0}}$ is the Hubble distance (where c is the speed of light and $\mathrm{H_0}$ the Hubble constant at present time), and 

\begin{equation}
    E(z) = \sqrt{\Omega_r (1+z)^4 + \Omega_m (1+z)^3 + \Omega_k (1+z)^2 + \Omega_{\Lambda}},
\end{equation}
with $\mathrm{\Omega_r, \Omega_m, \Omega_k, \Omega_{\Lambda}}$ representing radiation, matter, curvature, and cosmological constant densities, respectively. The comoving distance to the centre of the simulation box at $\mathrm{z=1}$ is $\mathrm{d_{C}(z=1) = 3363.07~Mpc}$. The comoving co-ordinates of galaxies in the simulation box are known and can be added/subtracted to/from the `snapshot redshift' (in the chosen line-or-sight direction) to account for their position relative to the centre. The cosmological redshift $\mathrm{z_{hub}}$ can then be obtained from the previously computed table of $\mathrm{d_{C}(z)}$ and $\mathrm{z}$.

The line-of-sight component of the physical peculiar velocity, $\mathrm{v_{pec}}$, can be straightforwardly computed for all galaxies in the simulation box. Since $\mathrm{v_{pec} << c}$, redshift and velocity are related by: $\mathrm{z_{pec} \equiv v_{pec}/c}$. Finally, the observed redshift, $\mathrm{z_{obs}}$, can be computed via:

\begin{equation}
    (1+z_{obs}) = (1+z_{hub})(1+z_{pec}).
\end{equation}

With observed redshifts computed for every galaxy in the simulation box, $\mathrm{|\Delta z|}$ can be computed for every cluster of interest, and members selected using the vB20 criterion.

To obtain the velocity dispersion from observed cluster member line-of-sight velocities we use the `gapper' algorithm \citep{Beers_etal_1990}, which has been successfully used on observed clusters by \citet{Eke_etal_2004} and \citet{Robotham_etal_2011}. Under this scheme, galaxy velocities are sorted in increasing order, then velocity dispersion is estimated by:

\begin{equation}
    \sigma = \frac{\sqrt{\pi}}{N(N-1)} \sum_{i=1}^{N-1} \omega_i g_i,
\end{equation}

\noindent where $\mathrm{\omega_i = i(N-i)}$ and $\mathrm{g_i = v_{i+1} - v_i}$; N is the number of galaxies in the group or cluster, and $\mathrm{v_i}$ is the $\mathrm{i^{th}}$ velocity from a list of the galaxy velocities (in one dimension), which has been sorted in ascending order.  Note that the velocity dispersion itself is not used in the selection, but is a useful quantity to compare the distribution of velocities of selected galaxies with.

In the following test, we choose clusters similar to those used by vdB20 but, since we want to demonstrate the wider impact of observational selection, we do not match the GOGREEN halo mass distribution yet. Instead, we select all clusters with $\mathrm{log_{10}(M_{200c}/M_{\odot}) \geq 14.0}$ so as to maximise the galaxy number counts. However, we \emph{do} impose the GOGREEN halo mass distribution when computing the GSMF and $\mathrm{f_q}$ for reasons outlined later. The rest of the analysis follows the main text, i.e. $\mathrm{30kpc}$ aperture measurements and SF-Q division outlined in Sec.~\ref{sec:halo_selection_fq_desig}.

Figure~\ref{fig:phase_space} shows the distribution of member galaxies under two different selection criteria (FOF and observational from vdB20), divided into three bins of host halo mass (rows), and plotted showing two types of cluster-centric radii (columns). The x-axis is semi-logarithmic to best show the full range in radii, with linear scale in the inner regions of the cluster (within $\mathrm{r_{200c}}$) and logarithmic scale outside to better show the galaxy distributions. A consequence of this is the artificial `pile-up' of galaxies just beyond $\mathrm{r_{200c}}$. This metric will test the effects of imposing an aperture limit on galaxy selection.
The y-axis shows the spread in cluster-centric velocity, taking the BCG as the centre and normalising by line-of-sight velocity dispersion. This metric will highlight any contaminants being introduced as a result of the generous line-of-sight velocity dispersion cut.

Examining the left-hand column of Figure~\ref{fig:phase_space} reveals that FOF member galaxies extend well beyond $\mathrm{r_{200c}}$ (represented by a dashed line). The most distant galaxies can be found as far as $\mathrm{4r_{200c}}$ from the cluster centre in projected space.
By contrast, the observational selection is truncated in two dimensions due to the fixed 1 Mpc aperture. Being a fixed aperture, it has a more pronounced effect on high-mass haloes as evidenced by the truncation moving to progressively lower values of $\mathrm{r_{2D}/r_{200}}$. For reference, the three halo mass bins have mean $\mathrm{\bar{r}_{200c} = [0.72, 0.86, 1.14]~Mpc}$, meaning that the $\mathrm{R=1~Mpc}$ cut lies slightly \emph{inside} of $\mathrm{r_{200c}}$ for the most massive clusters. This cut is conservative for GOGREEN and has the net effect of excluding member galaxies.

Staying on the left-hand column but turning our attention to the LoS velocity distribution relative to the BCG, we see that the observational selection exhibits a much greater scatter in this measure. FOF galaxies are confined within $\pm 3\mathrm{\sigma_{LOS,obs}}$, whereas observational galaxies are within $\pm 25\mathrm{\sigma_{LOS,obs}}$. Looking at the right-hand column reveals that these high-$\mathrm{\sigma_{LOS,obs}}$ galaxies originate at high $\mathrm{r_{3D}/r_{200}}$, far beyond the most distant FOF members. This is a result of projection effects and a rather generous LoS velocity cut. For reference, the mean velocity dispersion for the three halo mass bins, as estimated using FOF members, is $\mathrm{\bar{\sigma}_{LOS,Obs} = [920, 1020, 1348]~km/s}$, whereas the LoS velocity cut is $\mathrm{|\Delta v| = c|\Delta z| \sim 6000~km/s}$. This clearly indicates there will be some LoS contamination, as galaxies with such velocities cannot possibly be bound to the cluster.

For additional analysis, some diagnostic information is displayed at the bottom of each panel in the right column of Figure~\ref{fig:phase_space}. From this, one can see that the observational selection achieves sample purity of $\mathrm{\approx 50-80\%}$, a false positive fraction of $\mathrm{\approx 22-45\%}$, and a false negative fraction of $\mathrm{\approx 6-29\%}$ relative to FOF selection. Sample purity and the fraction of false negative members increase with increasing halo mass, while the fraction of false positives decreases quite rapidly. This is consistent with observational selection being too conservative and introducing a large number of false positive members for low-mass haloes, while for the most massive haloes it is more likely to exclude member galaxies rather than add contaminants (although the numbers are quite close and sampling relatively poor). 

Field galaxies are predominantly star-forming, while cluster galaxies are more likely to be quenched. Including a large number of field galaxies in the cluster sample, while excluding some of the cluster galaxies may lead to changes in the measured GSMFs and, subsequently, quenched fraction. However, many of the nearby field galaxies may actually belong to neighbouring groups and be undergoing pre-processing, making them quenched. This would undo some of the effects of field contaminants, negating the bias. Since contamination varies substantially with halo mass, it is important to match the halo mass distribution of the sample to which the comparison is being made (GOGREEN in our case). 

In Figure~\ref{fig:gsmf_diff} we plot the star-forming/quiescent GSMFs and quenched fractions for both selections, taking the ratio of the two estimates in the bottom row to highlight any differences.  By examining GSMFs in the left column, we see that the numbers of, both, star-forming and quenched galaxies are underestimated by the observational selection as a result of the $\mathrm{R=1~Mpc}$ and GOGREEN mean halo mass being relatively high. In particular, the underestimation is higher for star-forming galaxies ($\sim 25\%$) than for their quiescent ($5 - 10\%$) counterparts, in agreement with our simple volume-based corrections above. We also checked that nearby galaxies undergoing pre-processing do not significantly affect the results, by further imposing a FoF selection onto the observational selection (i.e., we apply the observational selection to the true FoF members only). There was only marginal change to the curves, not enough to change the conclusions, i.e. correlated structure does not significantly affect the selection in this regime.

Since the star-forming GSMF is suppressed more than quiescent, the resulting quenched fraction (shown in the right panel of Fig.~\ref{fig:gsmf_diff}) is artificially elevated by $\sim 5\%$ relative to FOF selection. This is not enough to impact any of our conclusions (neither those of vdB20). 

We do note, however, that, while this observational selection does not bias the results for GOGREEN clusters in a conclusion-altering way, it would significantly affect a sample with lower halo masses. GOGREEN occupies the two higher halo mass bins in Fig.~\ref{fig:phase_space}, which are the least contaminated by the selection. This is not so in the lowest mass bin and a significant number of star-forming galaxies would be added to the sample: enough to alter the measured quenched fractions.

\section{Total stellar content of haloes}
\label{app:stellar_content}

\begin{figure*}%
    \centering
    \includegraphics[width=\textwidth]{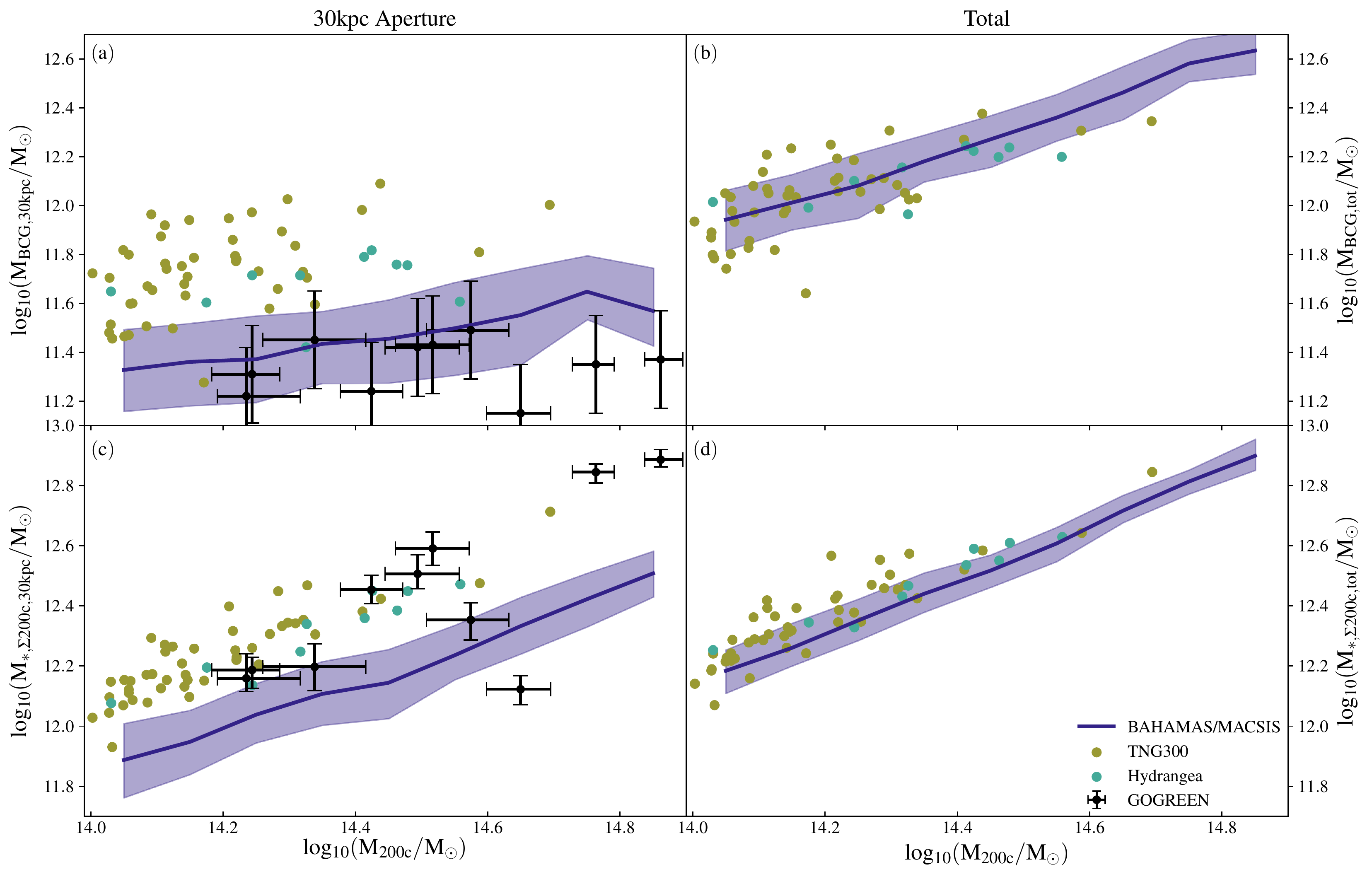}%
    \caption{\textit{Top row}: Stellar mass estimates of central galaxies with (left) and without (right) the 30~kpc aperture. Data from GOGREEN are shown as black points with error bars.  For all simulations, the total stellar masses of the central galaxies are higher than 30~kpc estimates, but BAHAMAS/MACSIS is particularly affected by it. This demonstrates that BAHAMAS/MACSIS centrals are significantly less compact than TNG300 and Hydrangea. Shaded regions indicate the 1$\mathrm{\sigma}$ scatter in the BAHAMAS/MACSIS sample. \textit{Bottom row}: total stellar mass within $\mathrm{R_{200c}}$ using the 30~kpc spherical aperture estimate as a function of halo $\mathrm{M_{200c}}$ (left) and corresponding total stellar mass estimate without using an aperture (right).  
    }
    \label{fig:mstar_bcg_halo}%
\end{figure*}

Here we examine Figure~\ref{fig:mstar_bcg_halo}, where we focus on the cluster halo mass range and show the distribution of masses in the simulations and data.  Panels (a) and (b) compare the central galaxy (BCG) masses to the halo masses (analogous to Figure~\ref{fig:mstar_mhalo_cent}) with and without the 30kpc spherical aperture.  Here the simulations are shown either as individual points (Hydrangea and TNG300) or as median value with 1$\sigma$ shaded region (BAHAMAS/MACSIS).  We see that most of the GOGREEN data are in reasonable agreement with the predictions of BAHAMAS/MACSIS, when stellar masses in the simulations are measured within a 30kpc aperture.  Total stellar mass ratios, shown in panel (b), are much higher - but the three simulations agree rather well.

In panels (c) and (d) we show the {\it total} stellar content, within a radius of $R_{200c}$.  Again we show the results considering masses computed within a 30kpc aperture (panel c), or using the total subhalo stellar mass (d).  The GOGREEN  measurements come from the  completeness-corrected sum of all stellar mass within $R_{\rm 200c}$, with an extrapolation to zero mass by fitting a Schechter function to each cluster, above its mass limit.  Uncertainties are estimated by bootstrap resampling, and including the uncertainty on the extrapolation due to uncertainties in the Schechter function fit parameters.
TNG300 and Hydrangea are in reasonable agreement on the power-law relation between stellar content and halo mass, with TNG300 having more haloes and, as a result, better samples the scatter in this regime. BAHAMAS/MACSIS is offset by $\approx0.2$~dex towards lower $\mathrm{M_{200c,*}}$ at all halo masses. These haloes are hosts to galaxies of lower stellar masses or are less compact (i.e. 30~kpc aperture cuts out a significant part of stellar mass) relative to the other two simulations.  This is confirmed in panel (d), which includes all star particles associated with the FOF group within $\mathrm{R_{200c}}$. Here, the offset is much smaller (below $1\sigma$ scatter) which suggests that BAHAMAS/MACSIS galaxies are substantially larger for the same host halo mass and more of the mass resides in the wings (intracluster light), instead. This is consistent with our findings for all central galaxies in the simulation.

\section{A closer look at the stellar mass function predictions of the different simulations}
\label{app:gsmfs}
Considering the total stellar mass functions shown in the left panel of Figure~\ref{fig:panel_gsmf}, there is a small excess abundance of about a factor of two at $\mathrm{log_{10}(M_*/M_{\odot}) < 10.5}$ in the BAHAMAS simulation, relative to the data. The same feature can be seen in figure~13 of  \citet{McCarthy_etal_2017}, which they argue is due to finite mass resolution, which limits the ability of low-mass galaxies to regulate their star formation rates (see the appendix of that study).

While BAHAMAS shows reasonable agreement for the cluster GSMF (right panel of Figure~\ref{fig:panel_gsmf}) at the lowest and highest masses, there is also a significant deficit near the knee of the mass function ($\mathrm{10.5 < \log_{10}(M_*/M_{\odot}) < 11.2}$).  One possible explanation for this feature is that tidal disruption of low-mass satellites may be overly efficient due to finite force resolution.  Alternatively (or perhaps additionally), it is known that density only-based substructure finders, such as \verb|SUBFIND|, can struggle to recover the full gravitationally-bound stellar mass of substructures \citep{Bahe_2021} and it is likely that this is more of an issue for comparatively lower resolution simulations.  Indeed, excess tidal disruption and/or substructure finder issues are consistent with Figure~\ref{fig:mstar_bcg_halo} in Appendix \ref{app:stellar_content}.  From that figure we conclude that the total stellar mass content summed over particles is approximately the same for the different simulations.  When limited to summing the mass of galaxies within 30 kpc apertures, however, BAHAMAS/MACSIS has lower mass compared to Hydrangea and TNG300 (see bottom left panel of Figure~\ref{fig:mstar_bcg_halo}).  Some of this difference is because Hydrangea and TNG300 have higher BCG masses within 30 kpc, but most of the effect is likely either because satellites are too efficiently stripped in BAHAMAS (their stellar masses are reduced within 30kpc) or they are destroyed altogether (or not detected by the substructure finder).  

Moving on to EAGLE/Hydrangea, the field AGNdT9 simulation box is too small to contain a representative number of galaxies above $\mathrm{log_{10}(M_*/M_{\odot}) = 10.6}$, which explains the premature decline of the field GSMF.  At low masses ($M_* < 10^{10.6}~\msun$), the amplitude is a reasonable match to the data but the slope is steeper, over-predicting the abundance of the lowest mass galaxies by a few tens of percent.  For the cluster comparison we can use the Hydrangea zoom simulations, using the same physics.  
We see similar behaviour to that in BAHAMAS: a reasonable match at the lowest masses, $\mathrm{log_{10}(M_*/M_{\odot}) \approx 10}$, but a lower abundance of galaxies compared with GOGREEN near the knee of the GSMF.  The discrepancy with GOGREEN at the knee of the GSMF is plausibly explained as a result of a slight mismatch in the halo mass selection for Hydrangea and GOGREEN, rather than the resolution-related issues discussed above for BAHAMAS/MACSIS.  Indeed, \citet{Ahad2021} found reasonably good agreement between Hydrangea and GOGREEN where they used a simple parametric scaling to account for the halo mass difference between Hydrangea and GOGREEN.  Using the BAHAMAS/MACSIS suite we have derived a factor of $\approx1.4$ to scale the GSMF from the Hydrangea sample to that of a sample with the GOGREEN mean halo mass.  We show that the scaled Hydrangea GSMF, shown in Figure~\ref{fig:panel_gsmf} with the dotted cyan curve, is in much better agreement with GOGREEN near the knee (i.e., consistent with \citealt{Ahad2021}).  Scaling the curves up by this factor does, however, exacerbate the differences with respect to GOGREEN at the very lowest and highest masses. In other words, while Hydrangea reproduces the amplitude of the cluster GSMF relatively well (once the difference in halo mass is accounted for), the shape of the predicted GSMF differs in detail from that observed in GOGREEN clusters.

Finally, turning to the GSMF predictions of TNG300, we see a reasonably good match to the field, for all but the highest stellar masses.  There is an excess of galaxies above $\mathrm{log_{10}(M_*/M_{\odot}) \approx 11.5}$ when compared to GOGREEN data, which is also apparent in Figure~\ref{fig:mstar_mhalo_cent}.  This gives the field GSMF a flattened appearance.  These very massive galaxies are highly likely to be BCGs at the centres of clusters which may indicate over-cooling of BCGs in the model\footnote{Using the scaled stellar masses from rTNG results in an improved agreement with the observations for the GSMF but does not significantly alter the main quenching results presented below.}.  A similar excess of massive galaxies is also seen in the cluster GSMF (right panel).   TNG300 also shows a similar deficit of galaxies near the knee with an excess at the highest masses, giving the GSMF a flat appearance.  However, as for the case of Hydrangea, most of the discrepancy at the knee is due to a slight mismatch in the mean halo mass of the TNG300 haloes and the GOGREEN systems.  Again, using BAHAMAS/MACSIS we derive a scaling factor to scale the TNG300 GSMF (dotted curve).  Similar to Hydrangea, the issue at the highest masses is worsened somewhat by this rescaling.  There is also a notable difference at $\mathrm{log_{10}(M_*/M_{\odot}) \lesssim 9.5}$ between TNG300 and EAGLE/Hydrangea but, without observational data, it is unclear which one is more realistic.

In Figure~\ref{fig:panel_gsmf_2} we showed the GSMFs in the data and simulations split according to star-forming vs.~quiescent status.  
The observed behaviour is broadly replicated by BAHAMAS in the field; at least in that quiescent galaxies are more abundant at the high-mass end (although not significantly so) and the two curves cross at $\mathrm{log_{10}(M_*/M_{\odot}) \approx 10.75}$. It overpredicts the abundance of star-forming galaxies by a factor of a few in the interval $\mathrm{11.0 < log_{10}(M_*/M_{\odot}) < 11.5}$, while also overpredicting the abundance of quiescent galaxies at the low-mass end. Neither of the features can be accounted for by quantitatively altering the definition of quiescence (Section \ref{sec:sims-sfr}),  as indicated by the shaded regions. For the cluster sample, the BAHAMAS/MACSIS mass function of star-forming galaxies is much flatter than in the field. This is in contrast with observations which show that the shape of the star-forming GSMF does not differ much between the field and cluster environments. The quiescent population is a better match but it shows the same features as what was seen for the total GSMF (largely because quiescent galaxies dominate the total population in this sample).  Again these differences cannot be accounted for by star-forming/quenched designation (shaded regions) nor by uncertainties associated with choosing ten GOGREEN-like haloes (error bars).

EAGLE reproduces the observed field trends reasonably well below stellar masses of
$\mathrm{log_{10}(M_*/M_{\odot})= 10.75}$. 
In the cluster population, Hydrangea performs comparatively well at matching the GOGREEN measurements.  To within the uncertainty induced by the SFMS offset, it matches the star-forming GSMF. However, it does not perform quite so well on the quiescent population: Hydrangea shows a steep increase in low mass ($\mathrm{log_{10}(M_*/M_{\odot}) \lesssim 9.75}$) quiescent galaxies, with a steady power-law decline towards higher stellar masses. It underestimates the abundance of intermediate-mass quiescent galaxies, such that there is a complete absence of the `knee' feature.
This behaviour at $z\approx1$ is in contrast to the behaviour of these simulations at the present day.  In particular, \citet{Bahe_etal_2017_hydrangea} showed that the Hydrangea quenched fractions of satellites (typically $0.8$ but with a mild stellar- and host- mass dependence) in massive clusters at $z\approx0$ exceeded that seen in the observations of \citet{Wetzel_etal_2012}, implying a larger-than-observed abundance of quiescent cluster galaxies at the present day (see figure~6 of that study). Lowest stellar masses aside, Hydrangea underestimates or, at best, matches the observed GOGREEN cluster quenched fractions. Evidently, the role of environment evolves substantially between $z\approx0$ and the present day in these simulations.

Finally, TNG300 shows qualitatively the best match to the observations in the field. 
As noted in the main text, the predictions are more sensitive to the definition of quiescence than the other simulation.
In addition, the abundance of quenched galaxies declines abruptly below $\mathrm{log_{10}(M_*/M_{\odot}) \approx 10.6}$, which is likely linked to the transition from a very effective (at quenching) low accretion-rate mode of AGN feedback at higher masses to a regime where stellar feedback and high accretion-rate AGN are less effective at quenching galaxies (see figure~3 and discussion in \citealt{Donnari_etal_2019}).
In the cluster sample, TNG300 is similar to BAHAMAS/MACSIS in that it does a reasonably good job of matching the quiescent GSMF in abundance and shape, albeit missing the knee.  The mass function for the star-forming galaxies is, on the other hand, typically a factor of 2 to 3 lower in amplitude than observed in GOGREEN clusters.
\section*{Affiliations}
$^{1}$Astrophysics Research Institute, Liverpool John Moores University, Liverpool, L3 5RF, UK\\
$^{2}$The University of Liverpool, Liverpool, L69 3BX, UK\\
$^{3}$The Cockcroft Institute of Accelerator Science and Technology, Warrington, WA4 4AD, UK\\
$^{4}$Department of Physics and Astronomy, University of Waterloo, Waterloo,
ON N2L 3G1, Canada\\
$^{5}$Waterloo Centre for Astrophysics, University of Waterloo, Waterloo, ON
N2L 3G1, Canada\\
$^{6}$Leiden Observatory, Leiden University, PO Box 9513, 2300 RA Leiden, The Netherlands\\
$^{7}$INAF - Osservatorio Astronomico di Trieste, via G.B. Tiepolo 11, 34143 Trieste, Italy\\
$^{8}$Laboratoire d'astrophysique, \'Ecole Polytechnique F\'ed\'erale de Lausanne (EPFL), 1290 Sauverny, Switzerland  \\
$^{9}$GEPI, Observatoire de Paris, Universit\'e PSL, CNRS, Place Jules Janssen, F-92190 Meudon, France \\
$^{10}$INAF - Osservatorio astronomico di Padova, Vicolo Osservatorio 5, IT-35122 Padova, Italy \\
$^{11}$Department of Physics \& Astronomy,
University of California, Irvine, 4129 Reines Hall, Irvine, CA 92697, USA\\
$^{12}$IFPU, Institute for Fundamental Physics of the Universe, via Beirut 2, 34014, Trieste, Italy.\\
$^{13}$Departamento de Ingenier\'ia Inform\'atica y Ciencias de la Computaci\'on, Universidad de Concepci\'on, Concepci\'on, Chile\\ 
$^{14}$Department of Physics and Astronomy, University of California, Riverside, 900 University Avenue, Riverside, CA 92521, USA \\
$^{15}$Departamento de Astronom\'ia, Facultad de Ciencias F\'isicas y Matem\'aticas, Universidad de Concepci\'on, Concepci\'on, Chile
$^{16}$Department of Physics, University of Helsinki, Gustaf H\"allstr\"omin katu 2a, FI-00014 Helsinki, Finland \\
$^{17}$Research School of Astronomy and Astrophysics, The Australian National University, ACT 2601, Australia\\
$^{18}$ Centre for Gravitational Astrophysics, College of Science, The Australian National University, ACT 2601, Australia\\
$^{19}$School of Physics and Astronomy, University of Birmingham, Edgbaston, Birmingham B15 2TT, England \\
$^{20}$Department of Physics and Astronomy, York University, 4700 Keele Street, Toronto, Ontario, ON MJ3 1P3, Canada \\
$^{21}$Departamento de Ciencias F\'isicas, Universidad Andres Bello, Fernandez Concha 700, Las Condes 7591538, Santiago, Regi\'on Metropolitana, Chile\\
$^{22}$European Space Agency (ESA), European Space Astronomy Centre, Villanueva de la Ca\~{n}ada, E-28691 Madrid, Spain\\
$^{23}$Centro de Estudios de Física del Cosmos de Aragón (CEFCA), Plaza San Juan 1, 44001 Teruel, Spain.\\
$^{24}$Department of Physics and Astronomy, The University of Kansas, 1251 Wescoe Hall Drive, Lawrence, KS 66045, USA\\
$^{25}$IRAP, Institut de Recherche en Astrophysique et Planétologie, Université de Toulouse, UPS-OMP, CNRS, CNES, 14 avenue E. Belin, F-31400 Toulouse, France
, Oxford Road, Manchester , UK\\
$^{26}$ European Southern Observatory, Karl-Schwarzschild-Str. 2, 85748 Garching, Germany\\
$^{27}$Department of Astronomy \& Astrophysics, University of Toronto, Toronto, Canada \\
$^{28}$Steward Observatory and Department of Astronomy, University of Arizona, 933 N. Cherry Ave. Tucson, AZ, 85721, USA \\

\bsp	
\label{lastpage}
\end{document}